\documentclass[pre,showpacs,preprint,floatfix]{revtex4}
\usepackage{amsmath}
\usepackage{epsfig}
\usepackage{color}

\begin{document}

\title{Percolation of particles on recursive lattices:\\
(I) a theoretical approach at the atomic level}
\author{Andrea Corsi\footnote{Present address: Dipartimento di Fisica, Universit\`a degli Studi di Milano, Milano, Italy}, P. D. Gujrati\footnote{Electronic address: pdg@physics.uakron.edu}}
\affiliation{The Department of Physics and The Department of Polymer Science, The University of Akron, Akron, Ohio, 44325, U.S.A.}
\date{\today }

\begin{abstract}
Powdered materials of sizes ranging from nanometers to microns are widely used in materials science and are carefully selected to enhance the performance of a matrix. Fillers have been used in order to improve properties, such as mechanical, rheological, electrical, magnetic, thermal, etc. of the host material. Changes in the shape and size of the filler particles are known to affect and, in some cases, magnify such enhancement. This effect is usually associated with an increased probability of formation of a percolating cluster of filler particles in the matrix. In this series of papers, we will consider lattice models. Previous model  calculations of percolation in polymeric systems generally did not take the possible difference between the size and shape of monomers and filler particles into account and usually neglected interactions or accounted for them in a crude fashion. In our approach, the original lattice is replaced by a recursive structure on which calculations are done exactly and interactions as well as size and shape disparities can be easily taken into account. Here, we introduce the recursive approach, describe how to derive the percolation threshold as a function of the various parameters of the problem and apply the new approach to the analysis of the effect of correlations among monodisperse particles on the percolation threshold of a system. In the second paper of the series, we tackle the issue of the effect of size and shape disparities of the particles on their percolation properties. In the last paper, we describe the effects due to the presence of a polymer matrix. 
\end{abstract}

\pacs{ }
\maketitle

\section{Introduction}

\label{intro} Percolation theory deals with the effects of varying the number of the interconnections among the particles of a given kind in a system: the latter are usually assumed to be distributed completely random \cite{Stauffer,Zallen}. The single most interesting aspect of percolation is the presence of a sharp phase transition at which a long-range ``connectivity'' suddenly appears in the system. Connectivity in this work will always mean that arising from the formation of a clusters due to nearest-neighbor physical proximity, which may or may not result in any chemical bonding. In the former case, the percolation is known as a bond percolation, and in the latter case, it is known as site percolation. In the former case percolation corresponds to the formation of a cluster of connected bonds where in the latter it corresponds to the formation of a cluster of sites. The \textit{percolation transition} occurs at the \textit{percolation threshold} as a consequence of a progressive increase in the interparticle connectedness or a network formation within the system. This change in connectedness can be a consequence of a change in the concentration of a species present in a system (site percolation) or of the occupation probability of the bonds describing interactions between particles in the system (bond percolation). This aspect makes the percolation model very appealing. It has been applied to countless problems 
in very different fields: the study of gelation of branched polymers, the description of polymer composites, the oil industry, the analysis of diffusion in porous media, the study of electrical and magnetic properties of disordered systems as well as epidemiology, just to name a few \cite{Grimmett}.

As a model used to describe disordered media, percolation is one of the simplest \cite{Grimmett}. Its attractions are manifold. It is easy to formulate but not unrealistic in its qualitative predictions for random media. Also, it has always been a playground for developing mathematical techniques and insights.

Finally, it is well endowed with beautiful conjectures that are easy to state but apparently rather hard to settle \cite{Grimmett}.

The percolation transition can be described easily in a lattice model by studying the dependence of the probability to have an infinite network of neighboring occupied sites present in the system, indicated as $p$, on the fraction of occupied sites on the lattice, $\phi $. A typical result is shown schematically in Figure \ref{Figure:percolation}. In the simplest possible case, the so-called \textit{random site percolation}, the occupation of one site of the lattice by the particle is completely independent from the occupation of the neighboring sites by the other particles.

The probability to have an infinite network in the system is identically zero as long as $\phi $ is less than a critical value, $\phi _{\text{c}}$, known as the \textit{percolation threshold}. As soon as the threshold value is reached, one or more networks of infinite size are formed and will be present for any larger value of the occupation probability \cite{Efros,Stauffer}. The existence of a perfectly sharp threshold at the percolation transition is only true for an infinitely large system, in what is known as the \textit{thermodynamic limit} \cite{Stauffer,Zallen}. For a finite system, repeated experiments will yield a spread of observed thresholds with values close to, but not identical to, $\phi_{\text{c}}$. The thresholds will be distributed in a critical region with a width inversely proportional to the number of sites present in the lattice under consideration. This width goes to zero as the size of the system becomes infinitely large \cite{Efros}.

It should be stressed that the percolation transition is not a thermal singularity: all the derivatives of the free-energy of the system taken with respect to the temperature are continuous. The transition has some remarkable similarity with a continuous transition, though. In particular, a divergence of the correlation length is observed at the percolation threshold, a feature that is characteristic of continuous thermal transitions. In the case of percolation, the correlation length of the system can be thought of as the size of the largest cluster present in the system. This quantity diverges as the percolation threshold is reached from below.

The discrete nature of the material at nanoscales becomes critical in determining the properties of the composite. A theory that averages over these length scales cannot do full justice to this discrete nature, since the averaging processes performed during the calculations are done on length scales much larger than the size of the nanoparticles. The discrete nature of the problem, however, can be captured in a lattice theory. For example, percolation of water in microemulsions refers to the percolation of water globules in a continuum oil background \cite{Safran-85}. The size of a water molecule is about 0.1 nm, and interparticle separation in liquid water is about 0.3 nm. A rough estimate of the size of oil molecules is about 2 nm, and of water globules is about 8nm \cite{Safran-85}. Thus, at the nanoscale that we are interested in, the water globules and oil should be both considered discrete and not continuum. As we will see below, there are several exact results available for lattice percolation so that we can compare our findings with them. In addition, the calculation of percolation probability that is one of our aims in this series of papers is easy to carry out in lattice models, but is not trivial in continuum percolation. This motivates our work based on a lattice model.

A composite that contains nanoparticles is characterized by nanoscopic inhomogeneities that cannot be described satisfactorily in a continuum model since the averaging processes performed during the calculations are done on length scales much larger than the size of the nanoparticles. 

Lattice models have been employed for many decades to study percolation problems. Random percolation was originally solved exactly in two limiting cases: in one dimension and on a Bethe lattice.

In one dimension, it is possible to have percolation only when $\phi =1$ \cite{Stauffer} and it is not possible to observe the side of the phase transition corresponding to $\phi >\phi _{\text{c}}$. Even if this problem seems somehow unusual, since only one side of the transition is accessible, it has been very important because of its similarities with the percolation process in higher dimension and also for its similarities with certain aggregation processes \cite{Kolb-85}.

Besides the one dimensional case, there is one other case that has been solved exactly in the early stages of the work on percolation, the Bethe lattice case. The Bethe lattice is a particular recursive structure, part of which is shown in Figure \ref{Figure:bethe lattice}, that is built up from smaller parts in a recursive fashion. It has infinite dimensionality since the ratio between the number of surface sites and bulk sites is a finite number equal to $(q-2)/q$ where $q$ is the coordination number of the lattice, that is the number of nearest neighbors of any site in the bulk of the lattice. The $\mathcal{B}_{i}$'s shown in the figure represent one
possible $i$th branch of the lattice. Every $\mathcal{B}_{i}$ is similar to $\mathcal{B}_{i+1}$ except that the latter is smaller in size than the former. Every $\mathcal{B}_{i}$ branches out into $(q-1)\mathcal{B}_{i+1}$ branches. The complete lattice is formed by connecting $q\mathcal{B}_{0}$ branches at the origin.

Since a Bethe lattice cannot be embedded in any finite-dimensional space, the results obtained by studying the percolation process on this lattice represent the mean-field behavior for percolation processes \cite{Zallen}. The very nature of a Bethe lattice is the reason why it is possible to solve many physical problems when they are defined on such a structure: different branches are independent and never cross each other. This powerful simplification is not available for regular lattices where the existence of closed loops allows for innumerable crossings. The absence of closed loops simplifies the problem enormously, since there is one single path that connects any two points, so that is possible to derive explicit equations for $p(\phi )$ and for $\phi _{\text{c}}$ \cite{Efros,Stauffer}. It is possible to prove that on a Bethe lattice the percolation threshold is $\phi_{\text{c}}=1/(q-1)$ \cite{Stauffer}. If the coordination number of the lattice becomes very large, the fraction of sites that need to be occupied in order to have percolation becomes smaller and tends to zero as the coordination number goes to infinity. In general it is possible to study two different kind of percolation processes: the \textit{bond} percolation and the \textit{site} percolation. The site and bond percolation processes are usually different so that the thresholds for these two processes can be very different but they are always the same on a Bethe lattice. In the following, we will always refer to site percolation. It should be kept in mind that the calculations carried out on a Bethe lattice also describe the bond percolation results.

The percolation problem has been also solved on other recursive structures that we will deal with in the following such as the triangular tree \cite{Essam} and a general $l$-sided Husimi tree \cite{PDG-JPA-01}.

In the past thirty years much progress has been achieved in the description of random percolation on many different lattices. But the exact results are few, except for the ones described above. Most of the known percolation thresholds are only numerical estimates obtained either numerically or analytically. Until recently, the only other known exact results for the bond percolation were those for the honeycomb, square, triangular and Kagome lattices and for the site percolation those on a triangular lattice, all the other results representing estimates of the percolation threshold value \cite{Deng_PRE_2005}. Recently, exact values for the percolation thresholds on Martini and Archimedean lattices were also obtained \cite{Scullard_1,Scullard_2}.

One of the main reasons for the development of the percolation theory and its consequent flourishing is to be found in the availability of computers that have allowed scientists to study and solve problems that would have otherwise been unmanageable. Since then, the percolation theory has became a part of the field of critical phenomena. Most of the effort has been put in the study of the critical exponents of the percolation transition, which are supposed to show a universal behavior and can therefore be described by the simplest model exhibiting a percolation threshold. But in designing composite materials it is important to understand the non-universal behavior as well. In particular, the behavior of the percolation threshold as a function of volume fraction, shape, orientation and correlations of the component phases remains a key problem in composite materials \cite{Bug-85,Mecke-02} that needs careful study.

Since the introduction of the percolation problem, Monte-Carlo simulations have been performed in order to be able to describe the process. Although many systems contain polydisperse fillers, most of the simulations have dealt with monomeric systems in which the size of all the species present in the system is the same \cite{Binder, Landau}. The simulations results presented so far only illustrate that the percolation threshold for discs of two different sizes depends on the ratio of the disks \cite{Phani-84,Dhar-97}. Recent results \cite{Mecke-02} show how the dependence of the percolation threshold on the composition of the system is stronger for composites in which the ratio between the size of the components is the largest.

In the case of Monte Carlo simulations, an important limitation is represented by the finite size scaling that the results have to go through. Monte-Carlo simulations are obtained by considering finite systems, sometimes very small, depending on the capacity of the computers used in the project, and the thermodynamic limit is obtained by studying the problem with systems of different sizes and trying to extrapolate from these results the behavior of an infinite system. In the case of simple problems, in which all the components have the same size, the scaling process is quite simple and, to some extent, well understood. When looking at complicated  systems, though, the scaling process is not at all trivial and can introduce artifacts that depend on the choice of the initial lattice and can hinder the physics of the problem \cite{Binder}.

The choice of the boundary conditions can be very important and influence the simulation. Usually, problems arise if the filler molecules are in a very dense system and the simulations become extremely time consuming. Most interactions between different species are not taken into account in order to make solving the problem feasible.

\section{Objective of the present research}

The fabrication of typical polymeric systems often requires the intermixing of several macromolecular fluids along with the addition of solid filler particles. The most classical example of inclusion of solid particles in a polymer matrix is represented by the addition of carbon black into a natural rubber matrix in order to improve its strength and processability. The use of powdered materials of sizes ranging from nanometers to microns is not limited to polymer composites but is common in many branches of materials
science. Fillers are carefully selected to enhance the performance of a matrix. Fillers have been used in order to improve mechanical \cite{Adams-93,Adams-Clay,Bigg-79,Chodak-99,Chodak-01,Ji-02,Kauly-96,Liang-99,Sharaf-01}, rheological \cite{Adams-93,Adams-Clay,Ghosh-00}, electrical \cite{Gokturk-93,Bigg-79,Chodak-99,Chodak-01,Lu-97,Mamunya-02,Michels-89,Mikrajuddin-99,Narkis-00,Ota-97,Roldughin-00,Strumpler-99,Wang-92,Yi-99}, magnetic \cite{Gokturk-93,Fiske-97} and thermal \cite{Lu-97,Bigg-79,Dietsche-99,Hansen-75,Mamunya-02,Bigg-95} properties of the host material. The elastic modulus of a filled resin, as an example, results from a \emph{complex interplay} between the properties of the individual components. There is enough experimental evidence to conclude that the properties of the composite are affected by a large number of parameters: the size, shape and distribution of the reinforcing particles as well as the interactions between the particles and the polymeric matrix.

A lattice model is promising, as discussed above, and can be used to capture these nanoscopic inhomogeneities. Previous lattice model calculations of percolation generally did not take into account the possible difference between the size of filler particles and usually neglected interactions. Such calculations have very important limitations represented by the use of random mixing approximation, the incompressibility of the model and the necessity for the monomers and voids to have the same size.

Our approach replaces the original lattice by a recursive lattice (RL), which is built up from its smaller parts in a recursive fashion, for example a Bethe lattice like the one shown in Figure \ref{Figure:bethe lattice}. A Husimi tree, see Figure \ref{Figure:Husimi_tree}, has also been used. The choice of the recursive lattice to be used is dictated by the model being investigated.

Choosing the proper coordination number, both lattices can locally approximate a square lattice of coordination number $q=4$. Replacing the original lattice with a RL is the only approximation that is made in our approach. We then solve the problem exactly on the RL using a recursive technique that will be explained in detail in the following. The fixed-point solutions of these recursion relations that relate the partial partition functions of different branches of the lattice describe the behavior in the bulk of the lattice. All interactions, including excluded-volume, can be treated exactly, as well as size and shape disparity between different
components of the system. Since our theory is exact, even though defined on a recursive lattice, and we do not approximate any thermodynamic potentials, as it is done in almost all other theories, the results will not violate thermodynamics.

Our recursive approach is a generalization of the conventional recursive approach used on a Bethe lattice; however, the generalization is not trivial in two important respects. In the first place, as we are dealing with correlated percolation, we need to worry about the interplay between thermal phase transitions and percolation, for which we need to be able to determine the free energies of various states and their impact on percolation. This is usually not a concern if one is merely looking for multiple solutions but not their free energies. The second aspect has to do with the presence of higher cycle fix-point solutions, and percolation in the corresponding states represented by these higher cycle fix-points. This latter aspect has never been studied before to the best of our knowledge. 
There is yet another important difference with the standard approach, which only deal with fix-point solutions, whereas our calculation can also be carried out for a \emph{finite system}, such as a system next to an infinite wall, where we need not approach a fixed-point of the recursion relations.

Our theory is given by the solution on a recursive lattice. The recursive lattices based on tree structures only capture weak correlations and are consequently not suitable, for example, for carrying out the calculation of critical exponents. However, the tree-structure based recursive theory has been applied by Gujrati and coworkers to study and describe many interesting properties of polymer systems including phase separation, critical points, loop formation in tree polymer gel, theta states, compressibility  effects, immiscibility loop, the Kauzmann paradox and the ideal glass transition \cite{PDG-Bowman-JCP-99,PDG-JCP1-98,PDG-JCP2-98,PDG-Chhajer-JCP-98,PDG-Corsi-PRL-01,PDG-Rane-Corsi-PRE-03,PDG-Corsi-PRE-03,Corsi-Masterthesis}. The predictions obtained using this recursive approach are more reliable than the conventional mean-field calculations in many different contexts, including spin glasses, linear and branched polymers and gauge theories \cite{PDG-PRL-95}. 

Recursive lattices such as a diamond hierarchical lattice that do not possess a tree structure can also be used to obtain exact results using recursive techniques. We have used such a lattice in our group \cite{PDG-PRA-90} to obtain non-classical exponents. These lattices have been invented for the sole purpose of carrying out renormalization group (RG) calculations near critical points in certain statistical mechanical models.\cite{Migdal-76, Kadanoff-76,Berker-79, Reynolds-79} The RG calculations become \emph{exact} on these lattices. As the justification for the RG calculation requires a diverging correlation length near a critical point, their
usefulness lies in predicting reliably the singular part of the free energy of the model, and therefore the critical exponents. They are not supposed to be reliable for the free energy itself except when they are exact. As our interest is not in critical exponents here, we will restrict ourselves to tree-structure based recursive lattices in the current series of papers. 

It should be pointed out that the RG calculations also give rise to recursion relations. Thus, from the point of view of the use of recursive techniques, our approach is not mathematically different from that in the RG calculations, except for the requirement of a diverging length scale in the case of critical phenomena where RG calculations find their usefulness. The technique has also been applied to study percolation of particles of the same size near the percolation threshold \cite{Harris-77, Kirkpatrick-77, Reynolds-77, PDG-JPA-80}. However, we are unaware of any such calculation where size disparity has been taken into
consideration, which is one of our aims. 

Here we introduce our approach and we describe how the recursive technique can be used to obtain the percolation threshold as well as the thermodynamics of systems made of particles. We apply this new approach first to the random percolation in very simple systems in order to test our method vs. known results.

We then study the effects of correlations on the percolation process in such systems and draw some conclusions.

In the following paper \cite{secondpaper}, we study the possible percolation of filler particles of different sizes and shapes on a recursive lattice. Although size disparity may in principle induce phase separation in this kind of system, recent rigorous calculations \cite{PDG-PRE-01} have proved that no phase separation in an athermal fully packed state of hard particle mixtures on a lattice is possible merely due to size disparity. 

Finally, in the third part of our study \cite{thirdpaper}, we describe the percolation of systems of different sized and shaped particles in the presence of a polymer matrix. To study these systems, we use a model for linear polymers that we have previously applied to the study of the melting and glass transition of linear polymers \cite{PDG-Corsi-PRL-01,PDG-Rane-Corsi-PRE-03,PDG-Corsi-PRE-03}.

\section{Random percolation on recursive structures}

In order to validate our approach, we initially apply the recursive method to simpler problems involving only small ``particles'' and no polymers. We defer to the next papers the investigation of systems made of monomeric particles (particles that occupy one lattice site) and stars or square particles on a Husimi lattice. Here, we focus on the percolation of particles of the same size on three different lattices: the Bethe lattice, the square Husimi lattice and the triangular Husimi lattice. The Husimi lattice is an infinite tree structure formed by a polygons. We will allow only two polygons to be attached at a given site in the Husimi lattice. The choice of the Bethe lattice, in particular, is very important since we can compare the results obtained with our recursive approach with the exact results obtained by Flory in his ground-breaking work \cite{Flory-41-I,Flory-41-II,Flory-41-III}.

Let us first consider the problem of random percolation of monomeric particles on a square Husimi lattice. A site is arbitrarily designated as the origin. It could be possible to designate a square as the \textit{center} of the lattice as well. Each of the squares that form the lattice has four sites. Out of these four sites, one is closer to the origin and the other three are not. The site close to the origin is called the base site and given an index $m$ and the other three sites an index $(m+1)$. We will often refer to the latter sites as the middle sites, the two that are the nearest neighbors of the base site, and the remaining site as the peak site of the square. The square containing the base site $m$ is labeled as the $m$th generation square, see figure \ref{Husimi labeling}. The origin of the lattice is labeled as the $m=0$ level and the level index $m$ increases as we move outwards from the origin to the periphery of the lattice. The level (that is, the generation) of each square corresponds to the number of squares that are present between the square and the origin of the tree. We will call the branch of the lattice that starts and lies above the $m$th level as the $\mathcal{C}_{m}$ branch of the tree. This branch has its base at the $m$th site. In some cases, the $\mathcal{C}_{m}$ branch may require additional indexes. The need for additional indexes can arise in the presence of a complex structured ground state for the system under investigation. In the following papers \cite{secondpaper,thirdpaper} we will see that it is sometimes necessary to discriminate between the middle sites and the peak site of the square. When this becomes necessary in the so called ``2-cycle'' scheme (see \cite{secondpaper, thirdpaper}), we will give the index $m+1$ to the middle sites and the index $m+2$ to the peak site. In this paper, we will just need the first labeling introduced above, which is usually referred to as ``1-cycle'' description of the lattice.

\subsection{Random percolation on a square Husimi lattice}

Let us consider an $m$th level square in the Husimi lattice, between the $m$th and the $(m+1)$th generations. As explained above, the square contains one site at the $m$th level and three sites at the $(m+1)$th level. If we are considering random percolation of monomeric species on this lattice, the site at the $m$th level can only be in one of two states: either it is occupied, this state being called $\text{S}$, or it is not occupied, the state being called $\bar{\text{S}}$. We are interested in the contribution of the $m$th branch $\mathcal{C}_{m}$ of the lattice, given that its base in a given state $\text{S}$ or $\bar{\text{S}},$ to the total partition function of the system. This contribution is called the partial partition function (PPF) of the branch. It is easy to see that the PPF only depends on the state of its base site. We denote the PPF of the $\mathcal{C}_{m}$ branch when the $m$th level site is in state $\alpha(=\text{S}, \bar{\text{S}})$ as $Z_{m}\left( \alpha \right) $. We then express this PPF in terms of the PPF's of the three sites at the $(m+1)$th level, so that the set of configurations that the system can adopt on the portion of the lattice that
lies above the $m$th level are described in terms of the configurations that can be adopted in the branches that lie above the $(m+1)$th level weighted by the local interactions in the $m$th square.

Following Gujrati \cite{PDG-PRL-95}, the recursion relations can always be written in the following form: 
\begin{equation}
Z_{m}\left( \alpha \right) =Tr\left( W\left( \alpha ,\left\{ \beta \right\}
\right) \right) \cdot \underset{\beta }{\prod }Z_{m+1}\left( \beta \right)
\end{equation}%
where $\beta $ represents the possible state of one of the three sites on the $(m+1)$th level, \{$\beta $\} represents the set of states $\beta $, and $W(\alpha ,\{\beta \})$ denotes the local Boltzmann weight due to the activities and the local interactions between $\alpha $ and $\beta $ states. 
In this particular problem, since we are interested in the random percolation process, we do not consider any interaction between different states. The amount of sites that are occupied in the lattice by the particles is controlled by an activity $\eta $ for a particle, which is related to the chemical potential per particle, $\mu $, through 
\begin{equation}
\eta =\exp \left( \beta \mu \right) .
\end{equation}%
Here, $\beta $ is the inverse temperature $1/T$ in the units of the Boltzmann constant. The total partition function for this problem is trivial and can be written as 
\begin{equation}
Z=\sum \eta ^{N_{\text{S}}}
\end{equation}%
where $N_{\text{S}}$ represents the number of sites that are occupied, and the sum is over distinct configurations of the particles on the lattice of $N$ sites.

When the site at the $m$th level is occupied by a particle, the state $\text{S}$, there are eight possible different configurations that the system can assume at the three $(m+1)$th level sites. These eight configurations are shown in figure \ref{Figure:Random configurations}.

In configuration (a), all the sites at the $(m+1)$th level are occupied. 
Therefore the contribution of this configuration to the partial partition function coming from this configuration is 
\begin{equation}
Z_{m+1}\left( \text{S}\right) Z_{m+1}\left( \text{S}\right) Z_{m+1}\left( 
\text{S}\right) .
\end{equation}

Configurations (b) to (d) all represent two occupied sites and an unoccupied one. Thus, the contribution from each one of these three configurations is 
\begin{equation}
Z_{m+1}\left( \text{S}\right) Z_{m+1}\left( \text{S}\right) Z_{m+1}\left( 
\bar{\text{S}}\right) .
\end{equation}

Configurations (e) to (g) all consist of one occupied site and two unoccupied ones. Thus, the contribution from each one of these three configurations is 
\begin{equation}
Z_{m+1}\left( \text{S}\right) Z_{m+1}\left( \bar{\text{S}}\right)
Z_{m+1}\left( \bar{\text{S}}\right) .
\end{equation}

Finally, configuration (h) presents all three sites unoccupied so that its contribution to the partition function is 
\begin{equation}
Z_{m+1}\left( \bar{\text{S}}\right) Z_{m+1}\left( \bar{\text{S}}\right)
Z_{m+1}\left( \bar{\text{S}}\right) .
\end{equation}

The recursion relation for $Z_{m}(\text{S})$, the partial partition function for the $m$th branch of the Husimi lattice, given that the $m$th level site is occupied, is therefore given by: 
\begin{eqnarray}
Z_{m}\left( \text{S}\right) &=&\eta \lbrack Z_{m+1}\left( \text{S}\right)
Z_{m+1}\left( \text{S}\right) Z_{m+1}\left( \text{S}\right) +3Z_{m+1}\left( 
\text{S}\right) Z_{m+1}\left( \text{S}\right) Z_{m+1}\left( \bar{\text{S}}%
\right)  \notag  \label{PPF for A} \\
&&+3Z_{m+1}\left( \text{S}\right) Z_{m+1}\left( \bar{\text{S}}\right)
Z_{m+1}\left( \bar{\text{S}} \right) +Z_{m+1}\left( \bar{\text{S}}\right)
Z_{m+1}\left( \bar{\text{S}}\right) Z_{m+1}\left( \bar{\text{S}}\right) ].
\end{eqnarray}
If the site at the $m$th level is unoccupied, we can write a similar expression since the possible configurations of the three sites at the $(m+1)$th level are the same as in the previous case: 
\begin{eqnarray}
Z_{m}\left( \bar{\text{S}}\right) &=&Z_{m+1}\left( \text{S}\right)
Z_{m+1}\left( \text{S}\right) Z_{m+1}\left( \text{S}\right) +3Z_{m+1}\left( 
\text{S}\right) Z_{m+1}\left( \text{S}\right) Z_{m+1}\left( \bar{\text{S}}%
\right)  \notag \\
&&+3Z_{m+1}\left( \text{S}\right) Z_{m+1}\left( \bar{\text{S}}\right)
Z_{m+1}\left( \bar{\text{S}} \right) +Z_{m+1}\left( \bar{\text{S}}\right)
Z_{m+1}\left( \bar{\text{S}}\right) Z_{m+1}\left( \bar{\text{S}}\right),
\end{eqnarray}
the only difference being that we do not have an activity for this state. When $n$ species are present in a system, we only need $n-1$ activities to control the composition of the system since the abundance of the remaining species is determined by the constraint that the sum of all the fractions associated with the different species add up to one.

We introduce the following ratios: 
\begin{equation}
x_{m}\left( \text{S}\right) =\frac{Z_{m}\left( \text{S}\right) }{Z_{m}\left( 
\text{S}\right) +Z_{m}\left( \bar{\text{S}}\right) },
\end{equation}
and 
\begin{equation}
x_{m}\left( \bar{\text{S}}\right) =\frac{Z_{m}\left( \bar{\text{S}}\right) }{%
Z_{m}\left( \text{S}\right) +Z_{m}\left( \bar{\text{S}}\right) }%
=1-x_{m}\left( \text{S}\right) .
\end{equation}

As one moves from a level that is infinitely far away from the origin towards the origin itself, the recursion relations will reach their fixed point (FP) solutions, $x_{m}\left( \alpha \right) \rightarrow x^{\ast}\left( \alpha \right).$ These fixed point solutions of the recursion relations describe the behavior in the interior of the Husimi lattice. Once the fixed point is reached, the value of $x_{m}\left( \alpha \right)$ becomes independent of $m$, so that we can write: 
\begin{eqnarray}
x_{m}\left( \text{S}\right) &=&x_{m+1}\left( \text{S}\right) =s \\
x_{m}\left( \bar{\text{S}}\right) &=&x_{m+1}\left( \bar{\text{S}}\right)
=1-s.
\end{eqnarray}

To determine the FP solutions, we introduce 
\begin{equation}
B_{m}=Z_{m}\left( \text{S}\right) +Z_{m}\left( \bar{\text{S}}\right) ,
\end{equation}
and express it in terms of the partial partition functions of $(m+1)$th
level branches: 
\begin{align}
B_{m} & =Z_{m}\left( \text{S}\right) +Z_{m}\left( \bar{\text{S}}\right) 
\notag \\
& =(1+\eta)[Z_{m+1}\left( \text{S}\right) Z_{m+1}\left( \text{S}\right)
Z_{m+1}\left( \text{S}\right) +3Z_{m+1}\left( \text{S}\right) Z_{m+1}\left( 
\text{S}\right) Z_{m+1}\left( \bar{\text{S} }\right)  \notag \\
& +3Z_{m+1}\left( \text{S}\right) Z_{m+1}\left( \bar{\text{S}}\right)
Z_{m+1}\left( \bar{\text{S}}\right) +Z_{m+1}\left( \bar{\text{S}}\right)
Z_{m+1}\left( \bar{\text{S}}\right) Z_{m+1}\left( \bar{\text{S}}\right) ] 
\notag \\
& \equiv B_{m+1}^{3}Q
\end{align}
where $B_{m+1}=Z_{m+1}\left( \text{S}\right) +Z_{m+1}\left( \bar{\text{S}}\right)$ according to the definition, and we have introduced a polynomial 
\begin{equation}
Q=\left( 1+\eta\right) \left[ s^{3}+3s^{2}\left( 1-s\right) +3s\left(
1-s\right) ^{2}+\left( 1-s\right) ^{3}\right] =\left( 1+\eta\right) .
\end{equation}

Using this polynomial and the recursion relations written above, we can write 
\begin{equation}
s=\frac{\eta}{1+\eta}
\end{equation}
so that the activity is directly related to the value of the ratio.

In order to determine the density of sites that are occupied as a function of the activity we need to write the total partition function of the system. The total partition function of the system at the $(m=0)$th level can be written considering all the possible configurations of the $(m=0)$th level site \cite{PDG-PRL-95}. The total tree is obtained by joining two $(m=0)$th level branches, one of which is shown in figure \ref{Figure:Husimi_tree}. In order to obtain this total partition function, we must consider all the 
configurations that the system can assume in the two squares that meet at the origin of the tree. This can seem trivial for a system in which there are only two possible states, the origin being either occupied or not occupied, but it will become very useful with more complicated problems.

In this case the total partition function of the system can be written as 
\begin{equation}
Z_{0}=\frac{Z_{0}\left( \text{S}\right) Z_{0}\left( \text{S}\right) }{\eta }%
+Z_{0}\left( \bar{\text{S}}\right) Z_{0}\left( \bar{\text{S}}\right) .
\end{equation}
The first term corresponds to an occupied site and the second one to an unoccupied one. The factor $\eta $\ in the denominator of the first term is necessary in order not to over count the activity. Each $Z_{0}\left(\text{S}\right) $\ term contains a factor $\eta $, see equation \ref{PPF for A}; thus by dividing by $\eta $\ we avoid counting the activity twice for the same monomer. The density of monomers or occupied sites is always defined as the ratio between the partition function describing configurations in which the origin is occupied and the total partition function of the system. So, its general expression is: 
\begin{equation}
\phi =\frac{Z_{0}\left( \text{S}\right) Z_{0}\left( \text{S}\right) /\eta }{%
Z_{0}},  \label{occupied_fraction}
\end{equation}
so that in this case the density of occupied sites is identical to the ratio 
$s$: 
\begin{equation}
\phi =\frac{\eta }{1+\eta }.  \label{simple_random}
\end{equation}%
In order to calculate the percolation probability we proceed following Gujrati \cite{PDG-JCP-93,PDG-Bowman-JCP-99} but we extend the calculation to the tree \cite{PDG-JPA-01}. We introduce the probability $R_{m}\leq 1$ that a site occupied at the $m$th generation is connected to a finite cluster of occupied sites at higher generations. Then $Z_{m}R_{m}$ denotes the contribution to the partial partition function of $\mathcal{C}_{m}$ due to all those configurations in which the site at the $m$th generation is connected to a finite cluster above the $m$th site. If we divide $Z_{m}R_{m}$ by $Z_{m},$\ we obtain a recursion relation for $R_{m}$. At the fixed-point solution, $R_{m}$\ approaches its fixed-point value $R$ given by the solution of an equation of the form 
\begin{equation}
R=\rho \left( R\right) \leq 1,
\end{equation}
where the right hand side of this equation is always in the form of a ratio of two polynomials in $R$ and it is also always positive since these two polynomials are always positive.

The percolation threshold geometrically corresponds to the slope of $\rho\left( R\right) $\ at $R=1$ being equal to one. The onset of percolation then corresponds to 
\begin{equation}
\left( \frac{d\rho}{dR}\right) _{R=1}=1.
\end{equation}

In the case of random percolation on a Husimi lattice, it is possible to write: 
\begin{align}
Z_{m}\left( \text{S}\right) R_{m}& =\eta \lbrack R_{m+1}^{3}Z_{m+1}\left( 
\text{S}\right) Z_{m+1}\left( \text{S}\right) Z_{m+1}\left( \text{S}\right) 
\notag \\
& +3R_{m+1}^{2}Z_{m+1}\left( \text{S}\right) Z_{m+1}\left( \text{S}\right)
Z_{m+1}\left( \bar{\text{S}}\right)   \notag \\
& +(2R_{m+1}+1)Z_{m+1}\left( \text{S}\right) Z_{m+1}\left( \bar{\text{S}}%
\right) Z_{m+1}\left( \bar{\text{S}}\right)   \notag \\
& +Z_{m+1}\left( \bar{\text{S}}\right) Z_{m+1}\left( \bar{\text{S}}\right)
Z_{m+1}\left( \bar{\text{S}}\right) ],
\end{align}%
so that 
\begin{equation}
R_{m}=\frac{%
[R_{m+1}^{3}s_{m+1}{}^{3}+3R_{m+1}^{2}s_{m+1}^{2}(1-s_{m+1})+(2R_{m+1}+1)s_{m+1}(1-s_{m+1})^{2}+(1-s_{m+1})^{3}]%
}{[s_{m+1}^{3}+3s_{m+1}^{2}(1-s_{m+1})+3s(1-s_{m+1})^{2}+(1-s_{m+1})^{3}]}.
\label{roforhusimi}
\end{equation}

The form of the equation above can be understood by looking at the eight possible configurations of the three $(m+1)$th level sites of the lattice when the $m$th level is occupied, as introduced in figure \ref{Figure:Random configurations}. In configuration (a), all the sites at the $(m+1)$th level are occupied. In order for the $m$th level site to be connected to a finite cluster of occupied sites, all the three occupied sites at level $(m+1)$ must be connected to finite clusters above that level. Thus, the contribution to $Z_{m}\left( \text{S}\right) R_{m}$ due to this particular configuration is $R_{m+1}^{3}Z_{m+1}\left( \text{S}\right) Z_{m+1}\left(\text{S}\right) Z_{m+1}\left( \text{S}\right) $, with a factor $R_{m+1}Z_{m+1}\left( \text{S}\right) $ corresponding to each one of the three considered sites. When considering configurations (b) to (d) in figure \ref{Figure:Random configurations}, we notice that two of the three upper sites are occupied. These two sites must be connected to finite clusters of occupied sites in order for the $m$th level site to be connected to a finite cluster of sites.The third site is empty and consequently it does not matter if it is connected to a finite or infinite network of occupied sites since the cluster is ``broken'' already at the site under investigation. Then the contribution to $R_{m}Z_{m}\left(\text{S}\right) $ associated to each one of these configurations is $R_{m+1} ^{2}Z_{m+1}\left( \text{S}\right) Z_{m+1}\left( \text{S}\right) Z_{m+1}\left(\bar{\text{S}}\right).$ When looking at configurations (e) to (g), we have to be particularly careful. In the case of configurations (e) and (f), the $m$th level site is neighboring one occupied middle site at the $(m+1)$th level. This middle site must be connected to a finite cluster of occupied sites in order for the $m$th level site to be connected to a finite cluster of sites. Then the contribution to $R_{m}Z_{m}\left( \text{S}\right) $ associated to each one of these configurations is $R_{m+1}Z_{m+1}\left( \text{S}\right) Z_{m+1}\left(\bar{\text{S}}\right) Z_{m+1}\left(\bar{\text{S}}\right).$ In the case of configuration (g), instead, the only occupied site at level $(m+1)$ is the
peak site. This site, though, is not a nearest neighbor of the base site at level $m$. Since the two intermediate sites are not occupied, the base site will be connected to a finite cluster of occupied sites with certainty. Thus, the contribution of this particular configuration to $R_{m}Z_{m}\left(\text{S}\right) $ is $Z_{m+1}\left( \text{S}\right) Z_{m+1}\left( \bar{\text{S}}\right) Z_{m+1}\left( \bar{\text{S}}\right) .$ Finally, in the case of configuration (h), since all the three $(m+1)$th level sites are not
occupied, the contribution of this configuration to $R_{m}Z_{m}\left( \text{S}\right) $ is $Z_{m+1}\left( \text{S}\right)Z_{m+1}\left( \bar{\text{S}}\right) Z_{m+1}\left( \bar{\text{S}}\right).$ By taking the sum of all these contributions, we obtain equation \ref{roforhusimi}.

At the fixed point, as explained above, $R$ does not depend on the level, and we can write 
\begin{equation}
R=\rho \left( R\right) =\frac{%
[R^{3}s^{3}+3R^{2}s^{2}(1-s)+(2R+1)s(1-s)^{2}+(1-s)^{3}]}{%
[s^{3}+3s^{2}(1-s)+3s(1-s)^{2}+(1-s)^{3}]}.
\end{equation}%
We can solve this equation recursively as well. We look for the smaller solution as a function of $s$ ($=\eta /(1+\eta )$).

Each term present in the denominator of the right hand side of the equation ($\rho\left( R\right)) $\ is also present in the numerator, except that in the numerator it is weighted by some proper power of $R$, so that $\rho\left(1\right) =1$. From this follows immediately that $R=1$ is always a solution of the equation. It is also obvious that $\rho\left( R\right) $ must lie above the straight line $y=R$ for large values of $R$ since it grows with some power of $R$ that is always larger than 1. Thus there are only two possible scenarios, shown in figure \ref{Figure: ro(R)} \cite{PDG-JCP-93}. If the situation is the one described in (a), then $R=1$ is the only physically acceptable solution and the solution u is unphysical since it corresponds to a value of $R$ larger than unity while $R$ is supposed to be a probability. If this is the case, then there is no percolation occurring in the system since the only physically acceptable solution is $R=1$. If the parameters of the problem are changed then u passes through $R=1$ and appears as the physical solution s at $R<1$, as shown in figure \ref{Figure: ro(R)}(b). In this case s represents the physically stable solution and percolation occurs \cite{PDG-JCP-93}. If the slope $\rho^{^{\prime}}$ at a fixed point is larger than unity it can be easily shown that the fixed point is unstable. So in figure \ref{Figure: ro(R)}, the fixed point that appears at larger values of $R$ is always
unstable and labeled u. As the slope changes, the fixed point at $R=1$ can become unstable.

Since $R$ represents the probability of the (occupied) site at the origin of the lattice being connected to a finite cluster of occupied sites on one half of the lattice, because of how $R_{m}$ is calculated, and since two half trees are joined together at the origin to form the complete tree, we define 
\begin{equation}
p=1-R^{2}
\end{equation}
as the percolation probability. This represents the probability for the occupied site at the origin to be connected to an infinite cluster of occupied sites on both sides of the origin. This is not the only possibility but this definition for the percolation probability seems to be the most natural one to be defined on a structure like this one since it checks the presence of an infinite cluster that spans the entire lattice on both sides of the origin. We can have looser or stronger definition of percolating clusters depending on the condition that we impose on our clusters. The behavior of $p$ as a function of $\phi$\ is shown in figure \ref{Figure:husimi_random_P}.

The percolation threshold is in this case: 
\begin{equation}
\phi _{c}\simeq 0.38195.
\end{equation}%
We observe that it could be easily shown that the bond percolation threshold for this lattice is different than the site percolation threshold that we are considering here by using equation (51) in \cite{PDG-JPA-01}, unlike in the case of Bethe lattices where, as explained above, site and bond percolation thresholds always coincide.

\subsection{Random percolation on a triangular Husimi lattice}

This kind of calculation can be carried out on any recursive lattice. If, for example, we consider the triangular Husimi lattice, shown in figure \ref{Figure:Husimi triangular}, the recursion relation between the partial partition functions at the $m$th and $(m+1)$th level has the form: 
\begin{equation}
Z_{m}\left( \text{S}\right) =\eta\lbrack Z_{m+1}\left( \text{S}\right)
Z_{m+1}\left( \text{S}\right) +2Z_{m+1}\left( \text{S}\right) Z_{m+1}\left( 
\bar{\text{S}}\right) +Z_{m+1}\left( \bar{\text{S}}\right) Z_{m+1}\left( 
\bar{\text{S}}\right) ].
\end{equation}

All the possible configurations corresponding to these terms are shown in figure \ref{Figure:Triangular configurations}.

Using the same argument used in the case of the square Husimi lattice, it is possible to calculate the density of occupied sites by using equation \ref{occupied_fraction}. It is easy to show that also in this case the expression for the fraction of occupied sites is the same as in the case of the square Husimi lattice (see equation \ref{simple_random}). In the case of the triangular Husimi lattice, the equation for the probability to have an occupied site at the $m$th level connected to a finite cluster of occupied sites at higher generations assumes the simple form: 
\begin{equation}
R=\frac{[R^{2}s^{2}+2Rs(1-s)+(1-s)^{2}]}{[s^{2}+2s(1-s)+(1-s)^{2}]},
\end{equation}
where $s$ is defined as in the case of the square lattice.

We look for the smaller solution as a function of $s$ (thus of $\eta$). The results of this calculation are shown in figure \ref%
{Figure:Husimi_random_triangular_P}, which plots the probability $p$ of percolation as a function of the fraction of occupied sites.

The percolation threshold in this case is: 
\begin{equation}
\phi _{\mathrm{c}}=0.5.
\end{equation}%
This result for the site percolation should be contrasted with the known result for bond percolation on a triangular Husimi lattice (see \cite{Essam,PDG-JPA-01})
\begin{equation}
\phi _{\mathrm{c}}\cong 0.40303.
\end{equation}

\subsection{Random percolation an a Bethe lattice}

Since we are also interested in studying the percolation process on the Bethe lattice, we have described the random percolation on this structure using our recursive approach. The results obtained in this case can further validate the efficiency and correctness of our approach since the result for the random percolation on a Bethe lattice is exactly known, as explained in the introduction. The recursion relation for $Z_{m}(\text{S})$, the partial partition function for the $m$th branch of the Bethe lattice given that the $%
m$th level site is occupied, is given by:  
\begin{equation*}
Z_{m}\left( \text{S}\right) =\eta\lbrack Z_{m+1}\left( \text{S}\right)
+Z_{m+1}\left( \bar{\text{S}}\right) ]^{q-1}, 
\end{equation*}
where $q$ is the coordination number of the lattice. This is a consequence of the nature of the Bethe lattice where all the $q-1$ branches above the $m$th level site are independent of each other and can either be occupied or not.

By defining $\text{S}$ as in the case of the Husimi lattice, we obtain 
\begin{equation}
s=\frac{\eta}{1+\eta}=\phi,
\end{equation}
and the equation for the probability to have an occupied site at the $m$th level connected to a finite cluster of occupied sites at higher generations
has the form:%
\begin{equation}
R=[Rs+1-s]^{q-1},
\end{equation}
which provides the percolation threshold%
\begin{equation}
\phi_{\mathrm{c}}=\frac{1}{q-1},
\end{equation}
in agreement with the known exact results.

\section{Correlated percolation for a system of same-size particles}

In order to be able to mimic the behavior of real systems, it is very important to study the effects of interactions on the percolation process. It is possible to extend the model described above to the case of correlated percolation. In the case of correlated percolation, the occupation of a site of the lattice is not random but depends strongly on the interactions with the nearest neighbors. Let us start by considering the case of the square Husimi lattice considered above. The total partition function of the system can be written as 
\begin{equation}
Z=\sum \eta ^{N_{\text{S}}}w^{N_{\text{C}}}
\end{equation}%
where here $N_{\text{S}}$ represents the number of sites that are occupied and $w$ is the Boltzmann weight for every nearest-neighbor contact between particles of different nature (or, in this case, between occupied and unoccupied sites) present in the system, $N_{\text{C}}$ being the number of such contacts. The weight $w$ is determined by the excess interaction energy  $\varepsilon $ as follows: 
\begin{equation}
w=\exp \left( -\beta \varepsilon \right) .
\end{equation}%
The excess energy $\varepsilon $ between two different species $1$ and $2$ is the excess 
\begin{equation}
\varepsilon \equiv e_{12}-{\frac12}(e_{11}+e_{22}),
\end{equation}
where $e_{ij}$ is the direct interaction energy between the species $i$ and $j$. In the following, we also replace $T/\varepsilon $ with $T$ so that the temperature is measured in units of interaction energy. This can always be done without affecting the physics of the problem. Only the temperature scale changes. The activity for the occupied state, $\eta $, is then related to $w$ through 
\begin{equation}
\eta =w^{-\mu }.
\end{equation}%
If $\mu $ is positive, then the system will lower its free energy by having more occupied sites and will consequently prefer such sites. If $\mu $ is negative, instead, the free energy of the system will be lower if the number of unoccupied sites is larger. The sign of $\mu $ determines the ground state of the system since at zero temperature entropy does not matter and the lowest energy state will be the most stable one.

Let us look back at the eight possible configurations inside the $m$th level squares shown in Figure \ref{Figure:Random configurations}. In the configuration (a), all the sites at the $(m+1)$th level are occupied. Therefore, the contribution of this configuration to the partial partition function coming from this configuration is 
\begin{equation}
Z_{m+1}\left( \text{S}\right) Z_{m+1}\left( \text{S}\right) Z_{m+1}\left( 
\text{S}\right) .
\end{equation}%
There is no extra weight we have to consider. The configurations (b) to (d) all present two occupied sites and an unoccupied one. The unoccupied site in all three configurations is neighboring two occupied sites; we need to take this into account by multiplying the partial partition function of this configuration by a factor $w^{2}$, one $w$ factor for each contact between an occupied and an unoccupied site. Thus, the contribution from each one of these three configurations is 
\begin{equation}
w^{2}Z_{m+1}\left( \text{S}\right) Z_{m+1}\left( \text{S}\right)
Z_{m+1}\left( \bar{\text{S}}\right) .
\end{equation}

The configurations (e) to (g) all consist of one occupied site and two unoccupied ones. In the configurations (e) and (f), the unoccupied sites are one at the peak site of the square and the other one at one of the middle sites. The total number of contacts between unoccupied and occupied sites is two. Thus, the contribution from each one of these two configurations to the partial partition function at the $m$th level is 
\begin{equation}
w^{2}Z_{m+1}\left( \text{S}\right) Z_{m+1}\left( \bar{\text{S}}\right)
Z_{m+1}\left( \bar{\text{S}}\right) .
\end{equation}

In the configuration (g), instead, the two unoccupied sites are at the middle sites and each of them has a contact with both the occupied site at the peak of the square and the occupied site at the base of the square, the total number of contacts between occupied and non occupied sites being equal to four. Then, the contribution of this configuration is: 
\begin{equation}
w^{4}Z_{m+1}\left( \text{S}\right) Z_{m+1}\left( \bar{\text{S}}\right)
Z_{m+1}\left( \bar{\text{S}}\right) .
\end{equation}

Finally, the configuration (h) presents all three sites unoccupied so that its contribution to the partial partition function is 
\begin{equation}
w^{2}Z_{m+1}\left( \bar{\text{S}}\right) Z_{m+1}\left( \bar{\text{S}}\right)
Z_{m+1}\left(\bar{\text{S}}\right) ,
\end{equation}%
since there are only two contacts between occupied and unoccupied sites.

The recursion relation for $Z_{m}($S$)$, the partial partition function for the $m$th branch of the Husimi lattice given that the $m$th level site is occupied, is, therefore, given by: 
\begin{align}
Z_{m}\left( \text{S}\right) & =\eta \lbrack Z_{m+1}\left( \text{S}\right)
Z_{m+1}\left( \text{S}\right) Z_{m+1}\left( \text{S}\right)  \notag \\
& +3w^{2}Z_{m+1}\left( \text{S}\right) Z_{m+1}\left( \text{S}\right)
Z_{m+1}\left( \bar{\text{S}}\right)  \notag \\
& +(2w^{2}+w^{4})Z_{m+1}\left( \text{S}\right) Z_{m+1}\left( \bar{\text{S}}%
\right) Z_{m+1}\left( \bar{\text{S}}\right)  \notag \\
& +w^{2}Z_{m+1}\left( \bar{\text{S}}\right) Z_{m+1}\left( \bar{\text{S}}%
\right) Z_{m+1}\left( \bar{\text{S}}\right) ].
\end{align}

The corresponding recursion relation for the unoccupied state is: 
\begin{align}
Z_{m}\left( \bar{\text{S}}\right) & =w^{2}Z_{m+1}\left( \text{S}\right)
Z_{m+1}\left( \text{S}\right) Z_{m+1}\left( \text{S}\right)  \notag \\
& +(2w^{2}+w^{4})Z_{m+1}\left( \text{S}\right) Z_{m+1}\left( \text{S}\right)
Z_{m+1}\left( \bar{\text{S}}\right)  \notag \\
& +3w^{2}Z_{m+1}\left( \text{S}\right) Z_{m+1}\left( \bar{\text{S}}\right)
Z_{m+1}\left( \bar{\text{S}}\right)  \notag \\
& +Z_{m+1}\left( \bar{\text{S}}\right) Z_{m+1}\left( \bar{\text{S}}\right)
Z_{m+1}\left( \bar{\text{S}}\right) .
\end{align}

By defining the proper ratios, as before, after introducing 
\begin{equation}
B_{m}=Z_{m}\left( \text{S}\right) +Z_{m}\left( \bar{\text{S}}\right) ,
\end{equation}%
it is possible to obtain the thermodynamics of the system, as done above.

When looking at the percolation probability, it is possible to write: 
\begin{align}
Z_{m}\left( \text{S}\right) R_{m}& =\eta \lbrack R_{m+1}^{3}Z_{m+1}\left( 
\text{S}\right) Z_{m+1}\left( \text{S}\right) Z_{m+1}\left( \text{S}\right) 
\notag \\
& +3w^{2}R_{m+1}^{2}Z_{m+1}\left( \text{S}\right) Z_{m+1}\left( \text{S}%
\right) Z_{m+1}\left( \bar{\text{S}}\right)  \notag \\
& +(2w^{2}R_{m+1}+w^{4})Z_{m+1}\left( \text{S}\right) Z_{m+1}\left( \bar{%
\text{S}}\right) Z_{m+1}\left( \bar{\text{S}}\right)  \notag \\
& +w^{2}Z_{m+1}\left( \bar{\text{S}}\right) Z_{m+1}\left( \bar{\text{S}}%
\right) Z_{m+1}\left( \bar{\text{S}}\right) ].
\end{align}

By expressing the equation above in terms of ratios and by considering the fixed point behavior, it is possible to obtain 
\begin{equation}
R=\frac{%
[R^{3}s^{3}+3w^{2}R^{2}s^{2}(1-s)+(2w^{2}R+w^{4})s(1-s)^{2}+w^{2}(1-s)^{3}]}{
[s^{3}+3w^{2}s^{2}(1-s)+(2w^{2}+w^{4})s(1-s)^{2}+w^{2}(1-s)^{3}]}.
\end{equation}

This expression is not as simple as the previous one for the completely random percolation, but can still be solved using a recursive approach. 

The sign of the chemical potential determines the ground state of the system at zero temperature: if the chemical potential is positive, the lattice is completely occupied in the ground state while it is completely empty for a negative chemical potential. In order to study the percolation process, it is necessary to consider negative values of the chemical potential. For positive values of $\mu $\ the occupied sites are the majority at any temperature and the system has always at least one percolating cluster of occupied sites. In the case of negative $\mu $, instead, the number of occupied sites grows as the temperature of the system increases, until the percolation threshold is reached and an infinite cluster appears. In this case, it is also possible to introduce the concept of critical temperature $T_{\text{c}}$, the temperature above which one or more infinite clusters of occupied sites appear in the system.

The results for the percolation probability as a function of the density of occupied sites are shown in figure \ref{Figure:Husimi_correlated_P_vs_phi}. It is important to observe that in this case the density of occupied sites does not have a simple expression like the one obtained for the random percolation case in equation \ref{simple_random}, but it must be explicitly calculated, as explained before, as the ratio between that part of the partition function that contains configurations in which the origin is occupied and the total partition function of the system (the density of occupied states is obtained, as always, from the general expression given in equation \ref{occupied_fraction}). The percolation threshold becomes larger as $\mu $ becomes more negative. As $\mu$ changes from 0 towards $-\infty $, the curve of $p$ as a function of $\phi $\ tends to the one obtained in the case of random percolation and shown in figure \ref{Figure:husimi_random_P}. This behavior is easy to understand if we remember that in our model the temperature is rescaled by the interaction energy between the particles and that the chemical potential is consequently a rescaled chemical potential $\mu \longrightarrow \mu \varepsilon $. The limit $\mu \longrightarrow -\infty $ corresponds to the limit $\varepsilon \longrightarrow 0$, that is, to the random percolation case. In this limit, we must keep $\mu \varepsilon  $\ fixed and equal to a negative value. Increasing the value of $\mu $\ from  $-\infty $ towards $0$ is analogous to increasing the interaction energy between the occupied and unoccupied states. Since the interaction is repulsive, as $\mu $\ becomes smaller in magnitude the configurations in which occupied sites are surrounded by other occupied sites become favorable thus lowering the percolation threshold for the system. As the magnitude of $\mu $ becomes smaller, occupied sites like more and more to be surrounded by other occupied sites so that the presence of these interactions leads to the formation of ``pathways'' that connect occupied sites and that lead to a decrease of the percolation threshold.

Figure \ref{Figure:Husimi_correlated_P_vs_T} shows the behavior of the percolation probability as a function of $T$. The percolation threshold increases as the chemical potential becomes larger in magnitude. We are not interested in the case of attractive interactions here, since the excess energies for physical interactions between particles of different kinds are usually positive. However, in the case of attractive excess interactions between different kind of particles (which can occur in the presence of specific interactions), the percolation threshold should increase with the strength of the interactions since the configurations with occupied sites surrounded by empty ones would be favored.

Figures \ref{Figure:Husimi_correlated_phic_vs_mu_summary} and \ref{Figure:Husimi_correlated_Tc_vs_mu_summary} summarize the dependence of the percolation threshold and the critical temperature on the value of $\mu$.

As expected, both the percolation threshold and the critical temperature increase as the value of the chemical potential becomes larger and larger in magnitude. The percolation threshold has a limiting value represented by the value obtained for the random case studied above, $\phi_{c} \simeq 0.38195.$ The critical temperature, instead, grows unbounded as the chemical potential becomes larger and larger in magnitude. The random percolation case corresponds to the absence of interactions, what is usually known as the
athermal limit. This limit corresponds to an infinitely large temperature, since at such temperature any interactions become negligible. So, an infinitely large and negative chemical potential corresponds to an infinite value for the critical temperature.

\section{Conclusions}

In this paper we have introduced a new recursive approach that we intend to use to study the percolation of particles of different size and shape that might be embedded in a polymer matrix. The approach allows us to take into account size and shape differences as well as to correctly account for the interactions between the different species present in the system.  All these aspects were not taken into consideration all together in the previous descriptions of percolation that we are aware of. 

We have introduced the machinery of our recursive approach in which the original lattice where the problem is defined is replaced by a recursive structure.

We have shown how it is possible to calculate the percolation threshold for a system of non interacting particles on three different recursive structures: the Bethe lattice, the Husimi lattice and the triangular Husimi lattice. We have shown how one can recursively obtain the percolation from the solution of a system of equations. We have also shown that where exact results are known, as in the case of the Bethe lattice, the recursive approach that we have introduced is able to reproduce such exact results. 

The effect of interactions on the percolation of a system of monodisperse particles on a Husimi lattice has also been analyzed to show how the recursive approach can take into account the interactions. The results show that an increase in the attractive interactions between the particles present on the lattice lowers the percolation threshold and the value of the critical temperature. This result agrees with many previous observations and conjectures \cite{Kikuchi-70,Coniglio-77,Kertesz-82a,Bug-85}.

Now that we have described the foundations of our approach, we move to the description of more complicated systems. In the following paper, we describe the percolation in systems made of particles with different sizes and shapes. Finally, in the third portion of this study, we describe the effects of the presence of a polymeric matrix on such polydisperse systems of particles. In particular, we will show how the percolation threshold decreases with the aspect ratio of the particles present in the system and how the stiffness of a polymer matrix as well as the nature of the interactions between fillers and matrix affect the percolation properties of the system.

\newpage
\bibliographystyle{unsrt}
\bibliography{bio}

\newpage

\begin{figure}
FIGURE CAPTIONS

\caption{Dependence of the probability to have an infinitely large cluster of occupied sites, $p$, on the fraction of occupied sites in the system, $\phi$. See text for details.}
\label{Figure:percolation}

\caption{Part of an infinite lattice known as Bethe lattice. The finite
portion shown in the figure is an example of a Cayley tree, and
contains no closed loops. $\mathcal{\bar{B}}_{n}$ indicates an \textit{n}th generation branch.}
\label{Figure:bethe lattice}

\caption{Portion of an infinite lattice known as Husimi tree.\hfill }
\label{Figure:Husimi_tree}

\caption{The structure of a Husimi tree.\hfill}
\label{Husimi labeling}

\caption{Possible configurations of the sites in the $m$th level square when
the base site is occupied. Filled circles represent occupied sites, empty
circles represent empty ones.}
\label{Figure:Random configurations}

\caption{Possible forms of $\rho(R)$ and fixed points. The stable point (s) is
the one relevant to the physics of the system. From \cite{PDG-JCP-93}.}
\label{Figure: ro(R)}

\caption{Random percolation on a square Husimi tree: probability that an
occupied site at the origin of the lattice is connected to an infinite cluster of
occupied sites as a function of the density of occupied sites.}
\label{Figure:husimi_random_P}

\caption{Part of an infinite lattice known as triangular Husimi tree.\hfill}
\label{Figure:Husimi triangular}

\caption{Possible configurations of the sites in the $m$th level triangle when
the base site is occupied.}
\label{Figure:Triangular configurations}

\caption{Random percolation on a triangular Husimi tree: probability that an
occupied site at the origin of the lattice is connected to an infinite cluster
of occupied sites spanning the entire lattice as a function of the density of
occupied sites.}
\label{Figure:Husimi_random_triangular_P}

\end{figure}
\newpage
\begin{figure}

\caption{Correlated percolation on a square Husimi tree: probability that an
occupied site at the origin of the lattice is connected to an infinite cluster
of occupied sites spanning the entire lattice as a function of the density of
occupied sites and of the chemical potential for an occupied site.}
\label{Figure:Husimi_correlated_P_vs_phi}

\caption{Correlated percolation on a square Husimi tree: probability that an
occupied site at the origin of the lattice is connected to an infinite cluster
of occupied sites spanning the entire lattice as a function of the temperature
of the system and of the chemical potential for an occupied site.}
\label{Figure:Husimi_correlated_P_vs_T}

\caption{Percolation threshold as a function of the chemical potential for the
percolation of monomeric species on a Husimi tree.}
\label{Figure:Husimi_correlated_phic_vs_mu_summary}

\caption{Critical temperature as a function of the chemical potential for the
percolation of monomeric species on a Husimi tree.}
\label{Figure:Husimi_correlated_Tc_vs_mu_summary}

\vspace{5cm}

\end{figure}

\newpage

\begin{figure}[tp]
\begin{center}
\epsfig{file=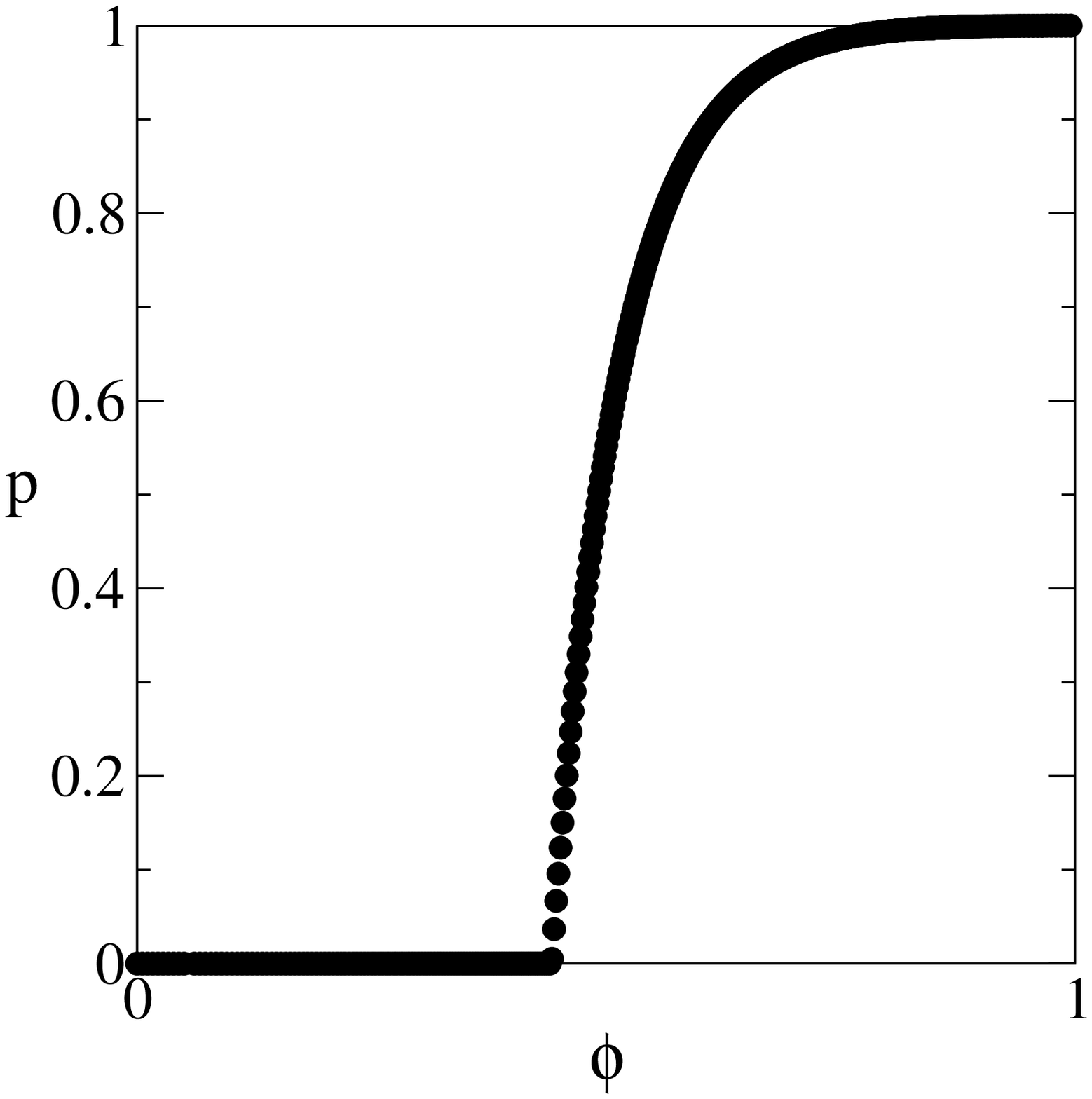,width=8.5cm,angle=270}
\end{center}
\par
FIG. 1 \vspace{5cm}
\end{figure}

\bigskip \bigskip \bigskip \newpage

\begin{figure}[tp]
\begin{center}
\vspace{1cm}
\epsfig{file=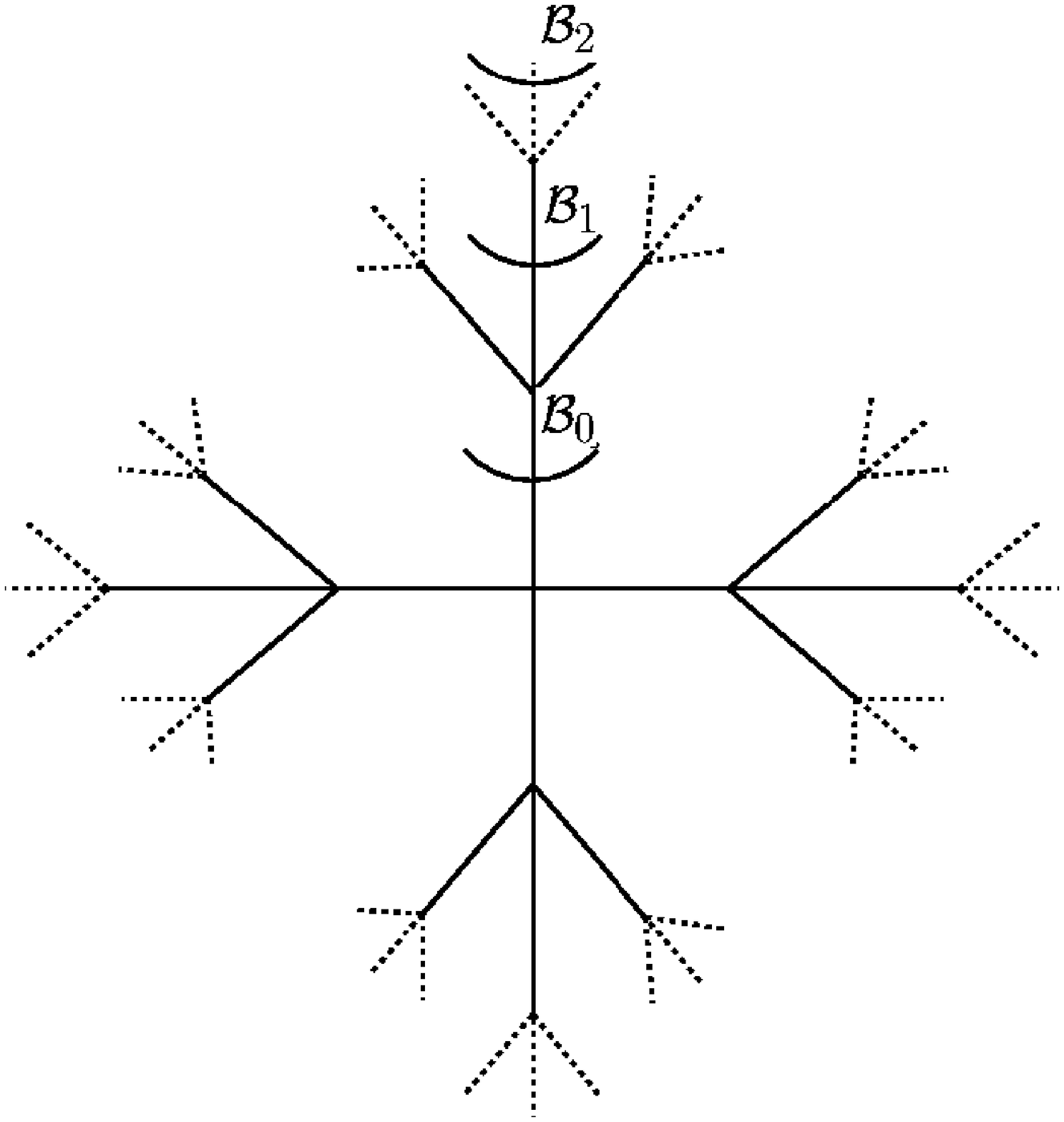,width=6.0cm}
\end{center}
\par
FIG. 2 \vspace{5cm}
\end{figure}

\newpage

\begin{figure}[tbp]
\begin{center}
\vspace{1cm}
\epsfig{file=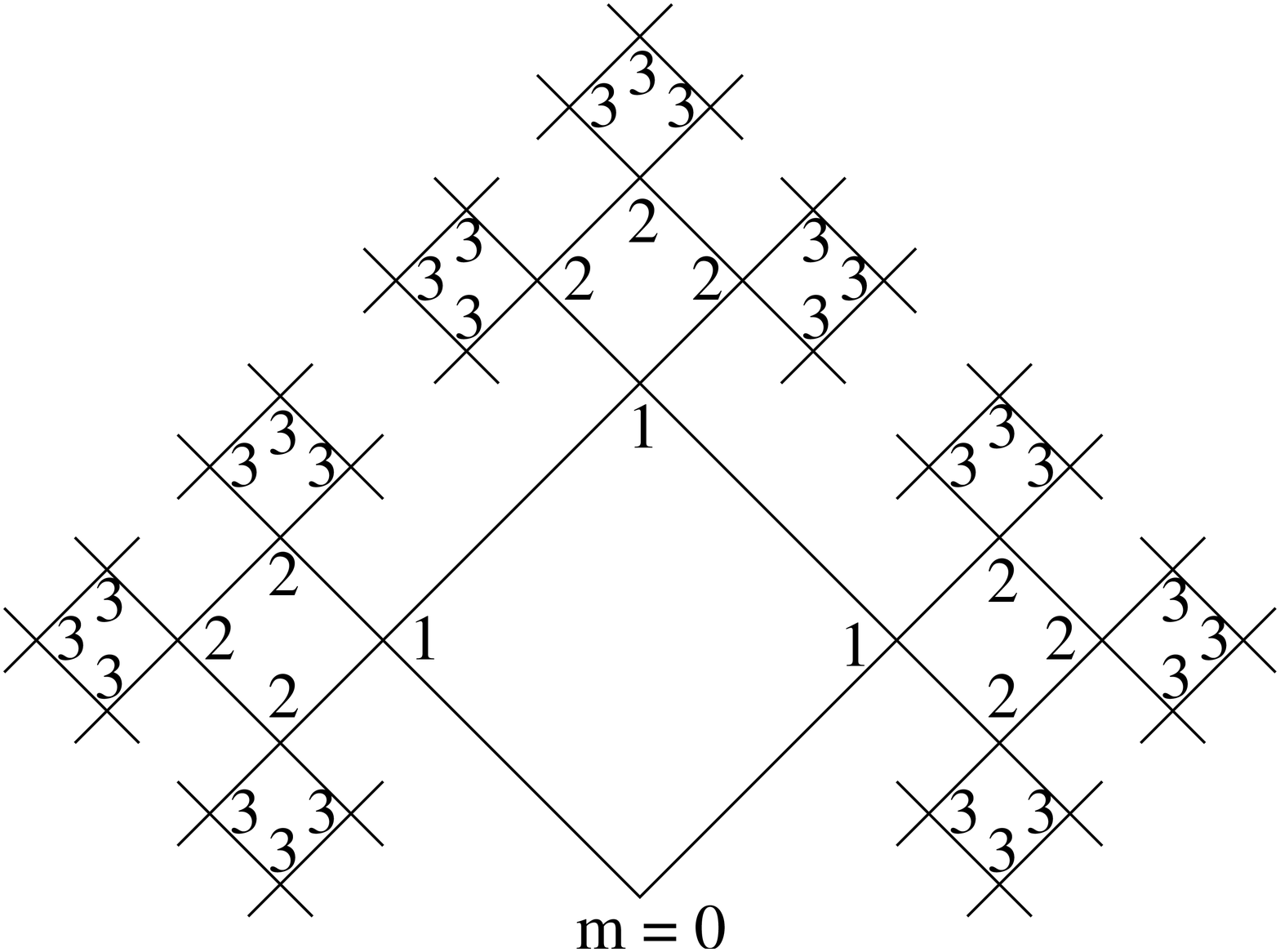,width=8.5cm}
\end{center}
\par
FIG. 3 \vspace{5cm}
\end{figure}

\newpage

\begin{figure}[p]
\begin{center}
\epsfig{file=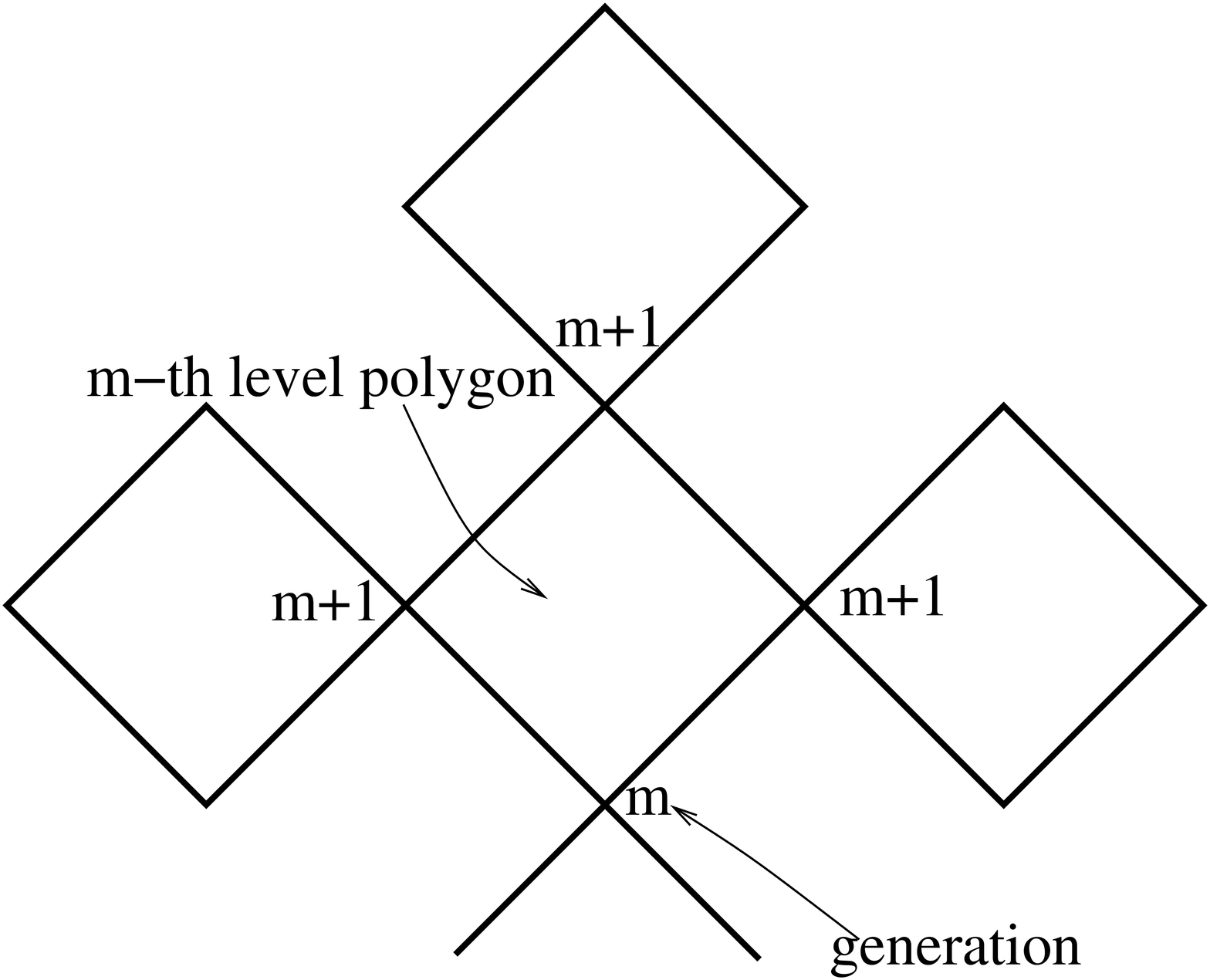,width=8.5cm}
\end{center}
\par
FIG. 4 \vspace{5cm}
\end{figure}

\newpage

\begin{figure}[p]
\begin{center}
\epsfig{file=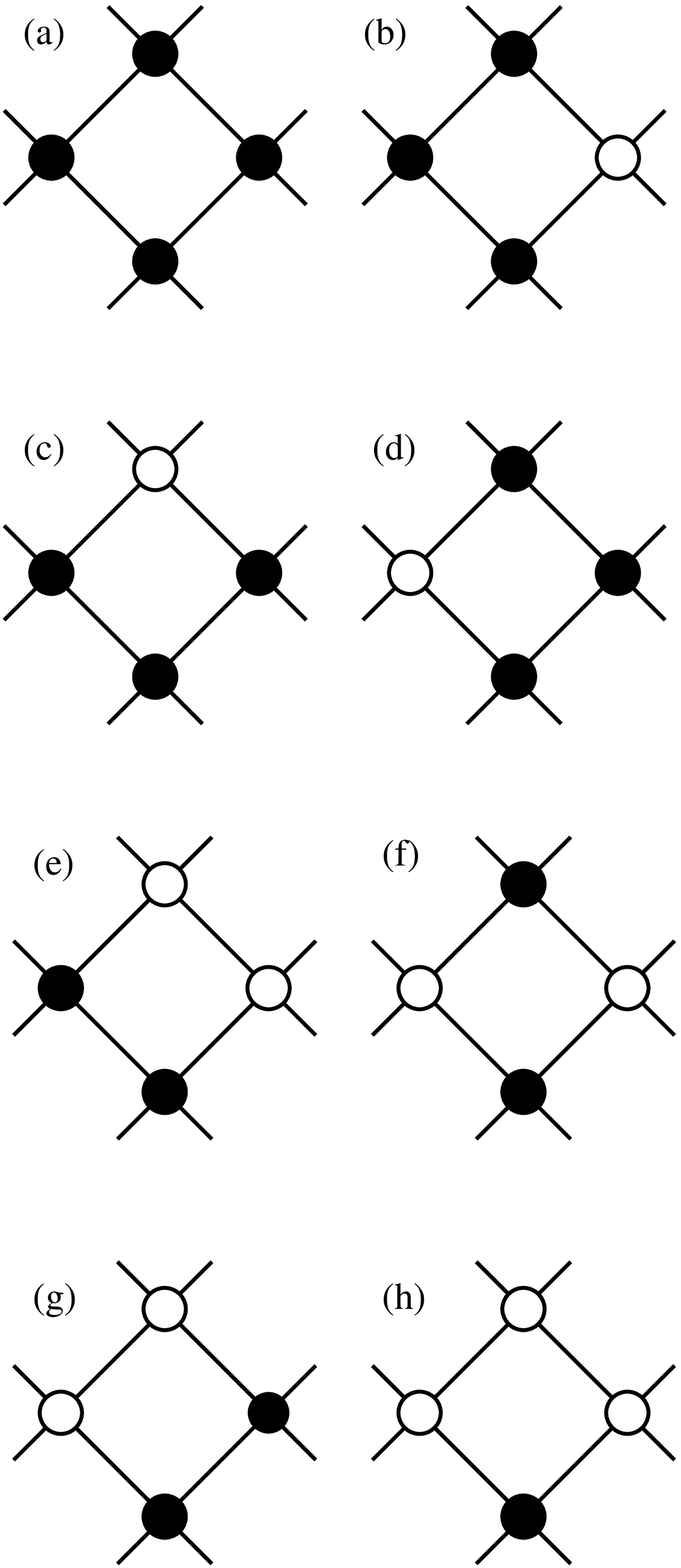,width=8.5cm}
\end{center}
\par
FIG. 5 \vspace{5cm}
\end{figure}

\newpage

\begin{figure}[p]
\begin{center}
\epsfig{file=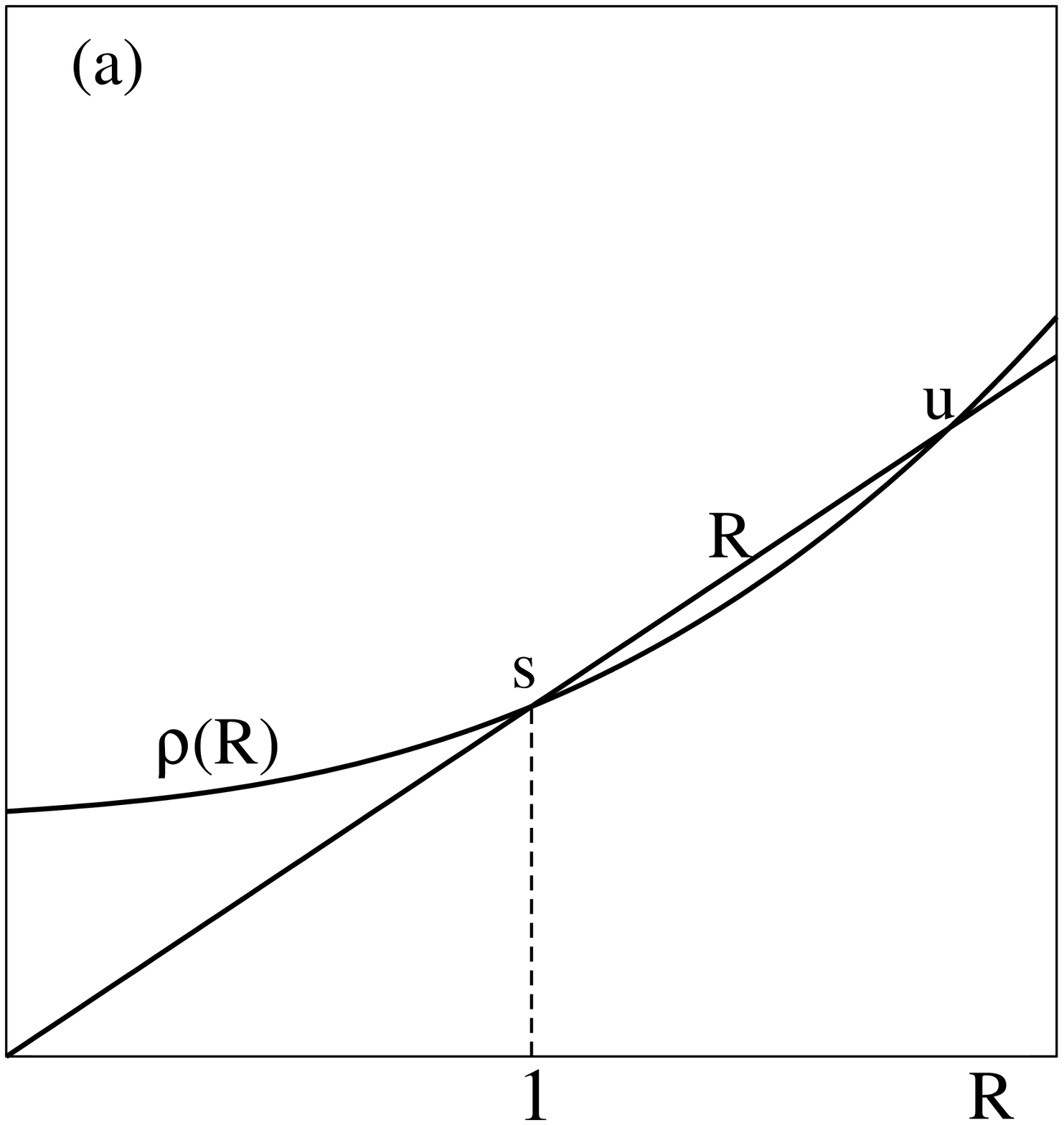,width=8.5cm,angle=270}
\\
\epsfig{file=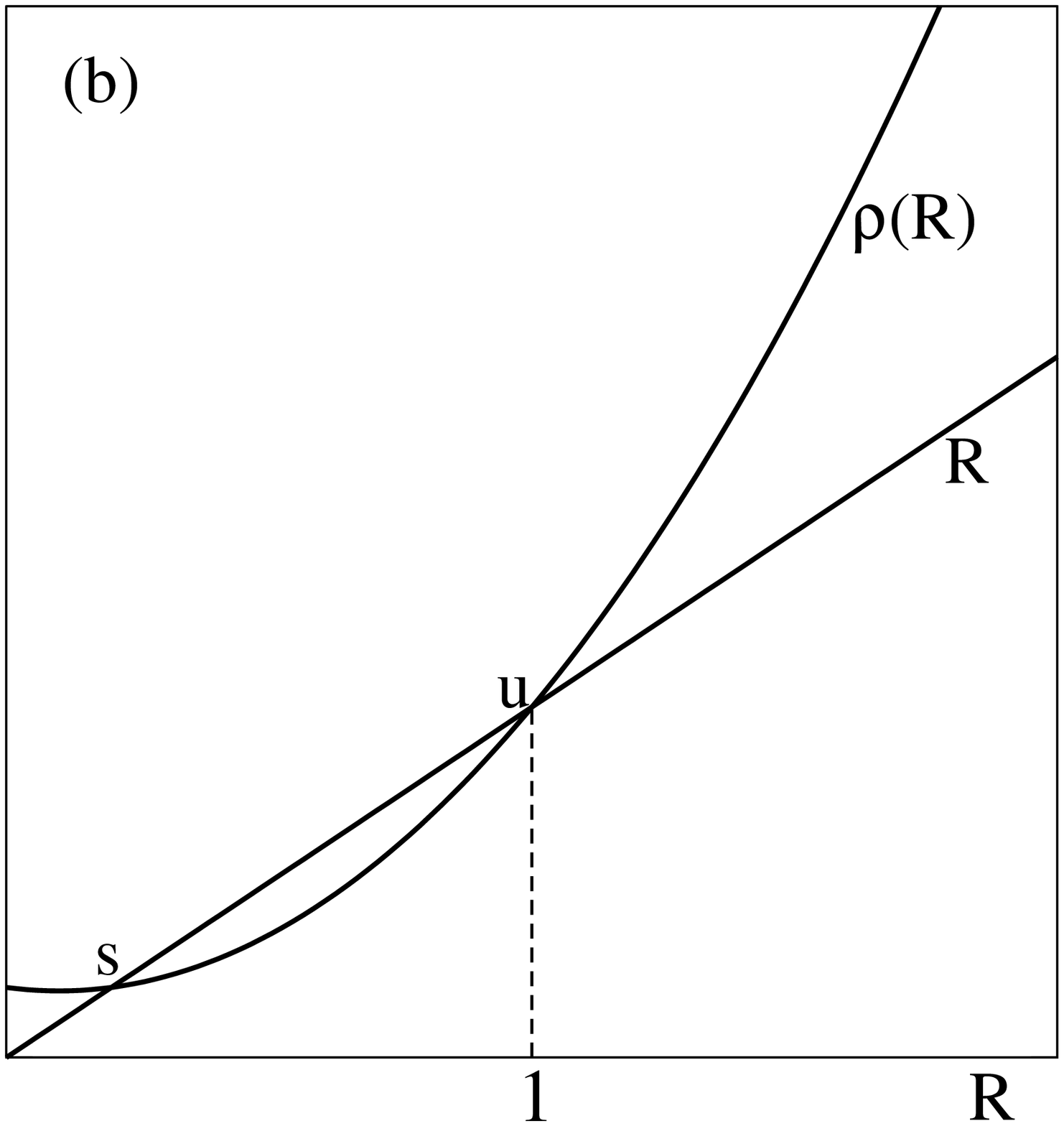,width=8.5cm,angle=270}
\end{center}
\par
FIG. 6 \vspace{5cm}
\end{figure}

\newpage

\begin{figure}[p]
\begin{center}
\epsfig{file=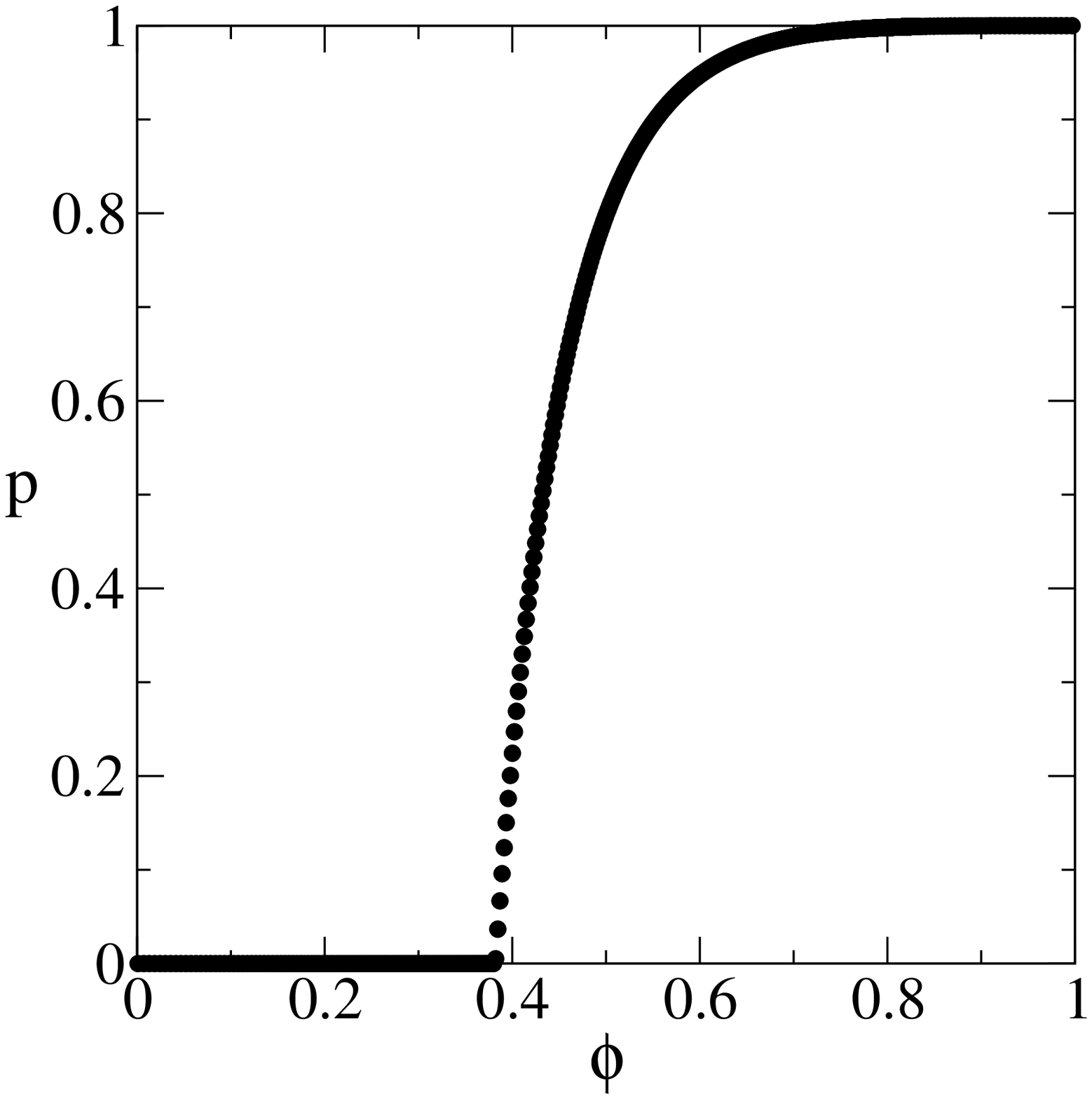,width=8.5cm,angle=270}
\end{center}
\par
FIG. 7 \vspace{5cm}
\end{figure}

\newpage

\begin{figure}[p]
\begin{center}
\epsfig{file=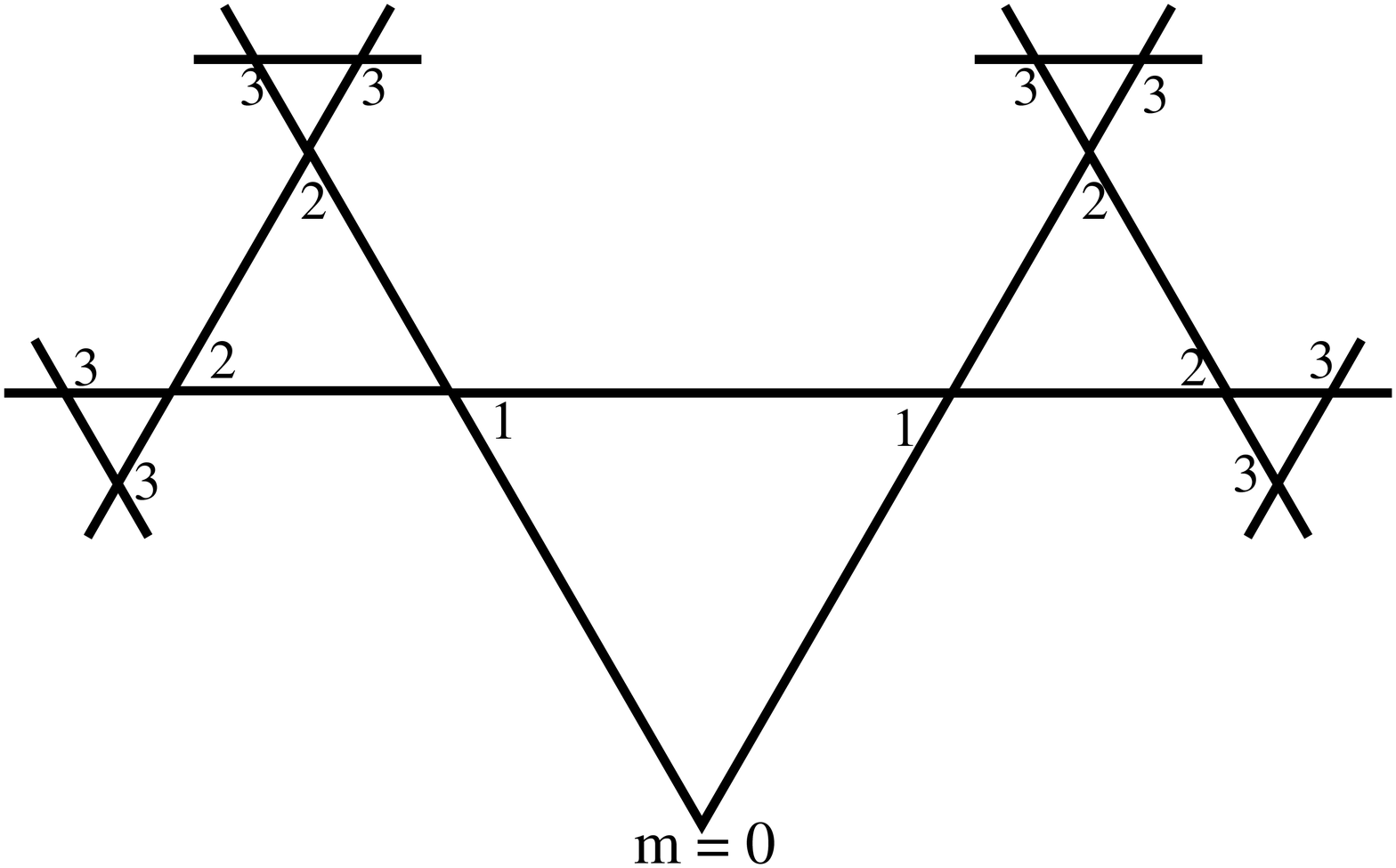,width=8.5cm}
\end{center}
\par
FIG. 8 \vspace{5cm}
\end{figure}

\newpage

\begin{figure}[p]
\begin{center}
\epsfig{file=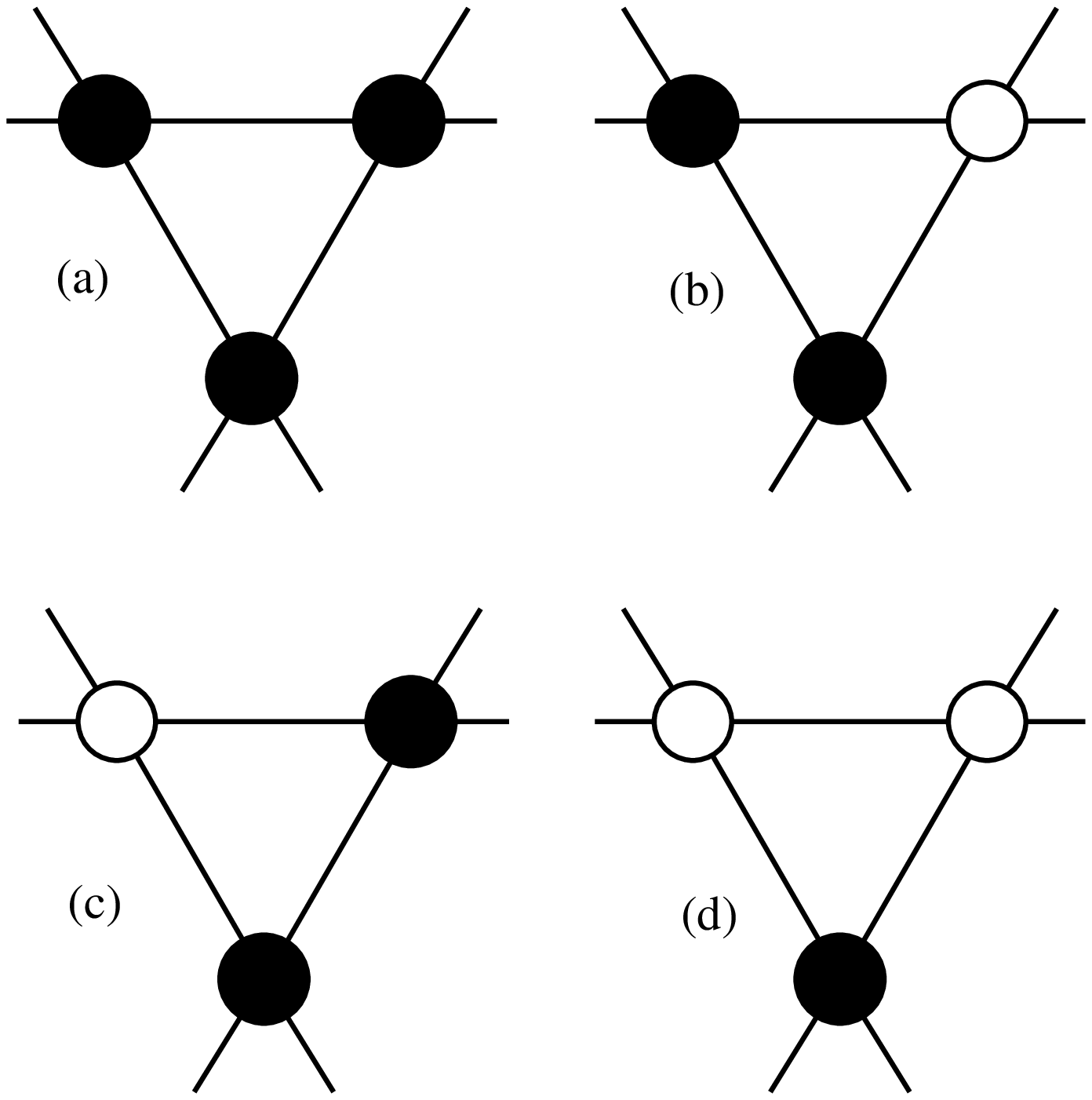,width=8.5cm}
\end{center}
\par
FIG. 9 \vspace{5cm}
\end{figure}

\newpage

\begin{figure}[p]
\begin{center}
\epsfig{file=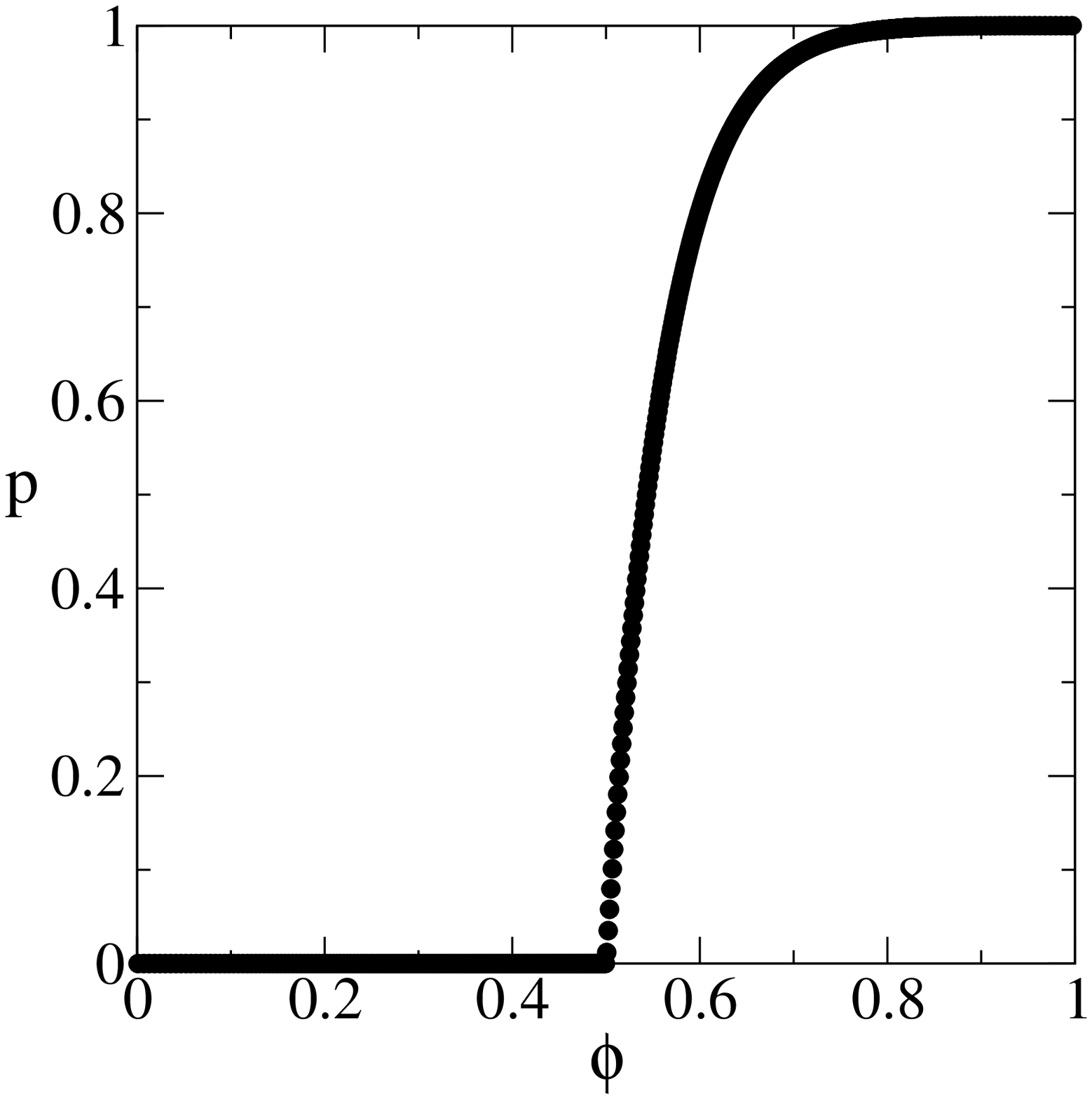,width=8.5cm,angle=270}
\end{center}
\par
FIG. 10 \vspace{5cm}
\end{figure}

\newpage

\begin{figure}[p]
\begin{center}
\epsfig{file=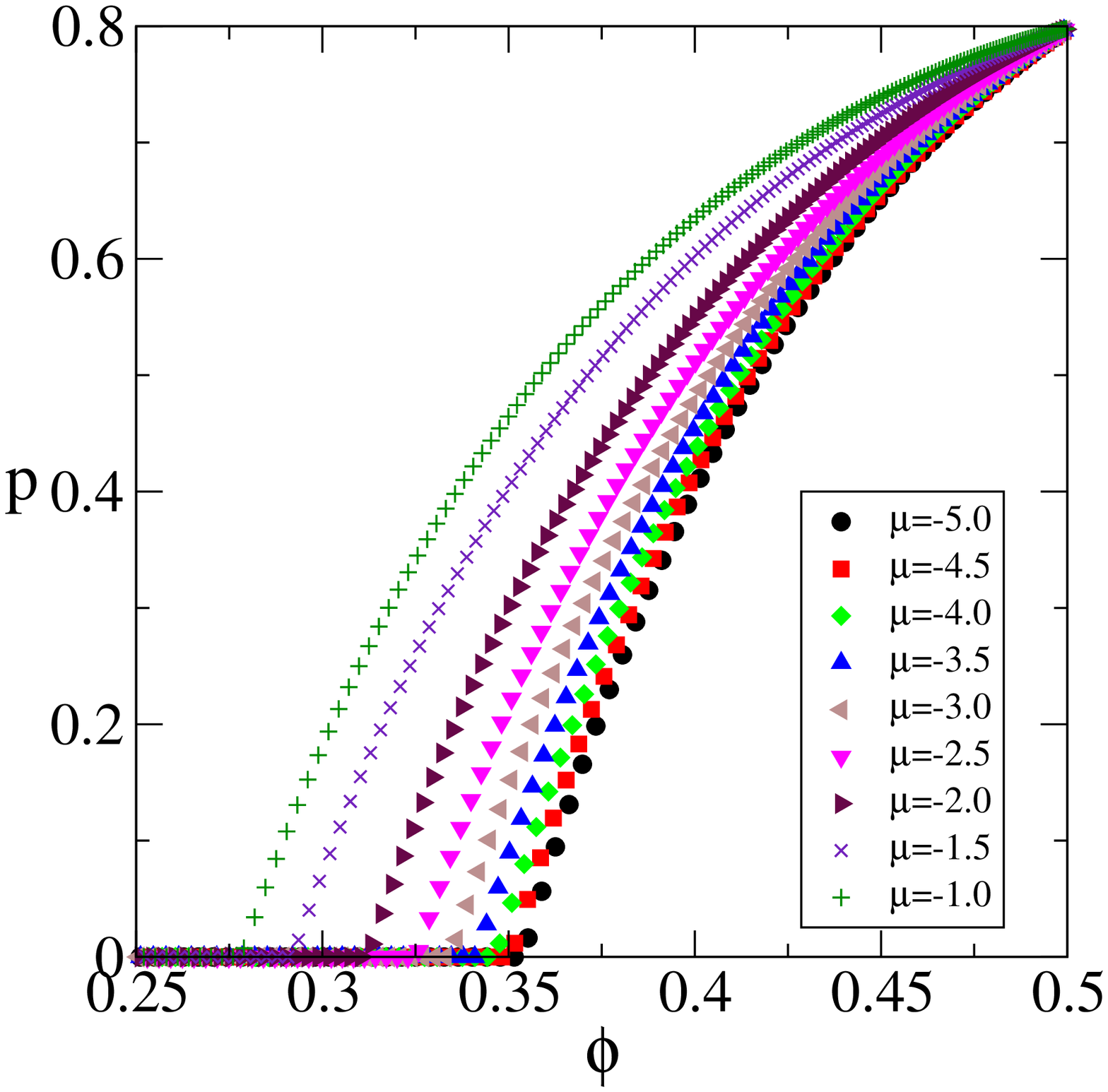,width=8.5cm,angle=270}
\end{center}
\par
FIG. 11 \vspace{5cm}
\end{figure}

\newpage

\begin{figure}[p]
\begin{center}
\epsfig{file=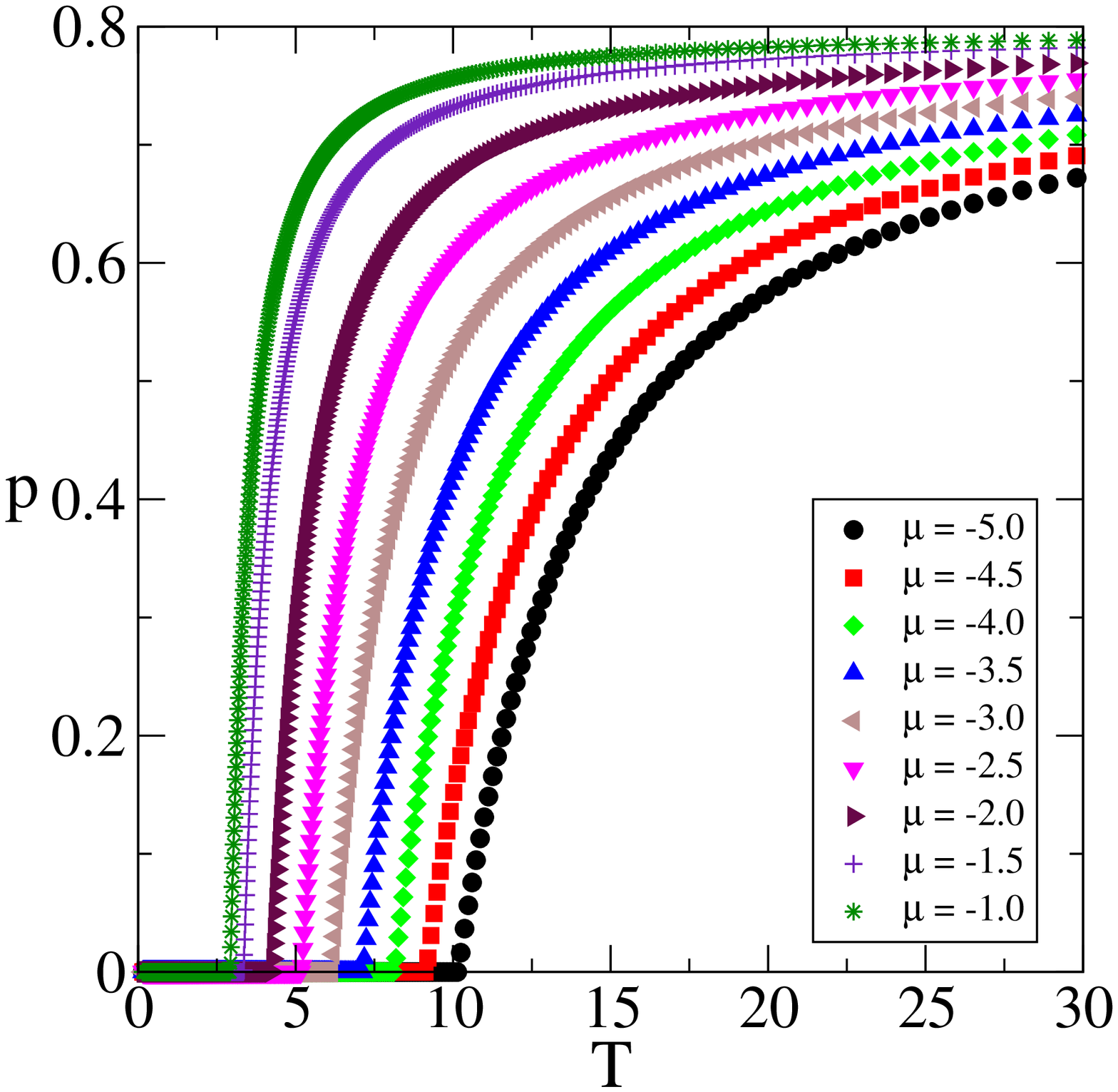,width=8.5cm,angle=270}
\end{center}
\par
FIG. 12 \vspace{5cm}
\end{figure}

\newpage

\begin{figure}[p]
\begin{center}
\epsfig{file=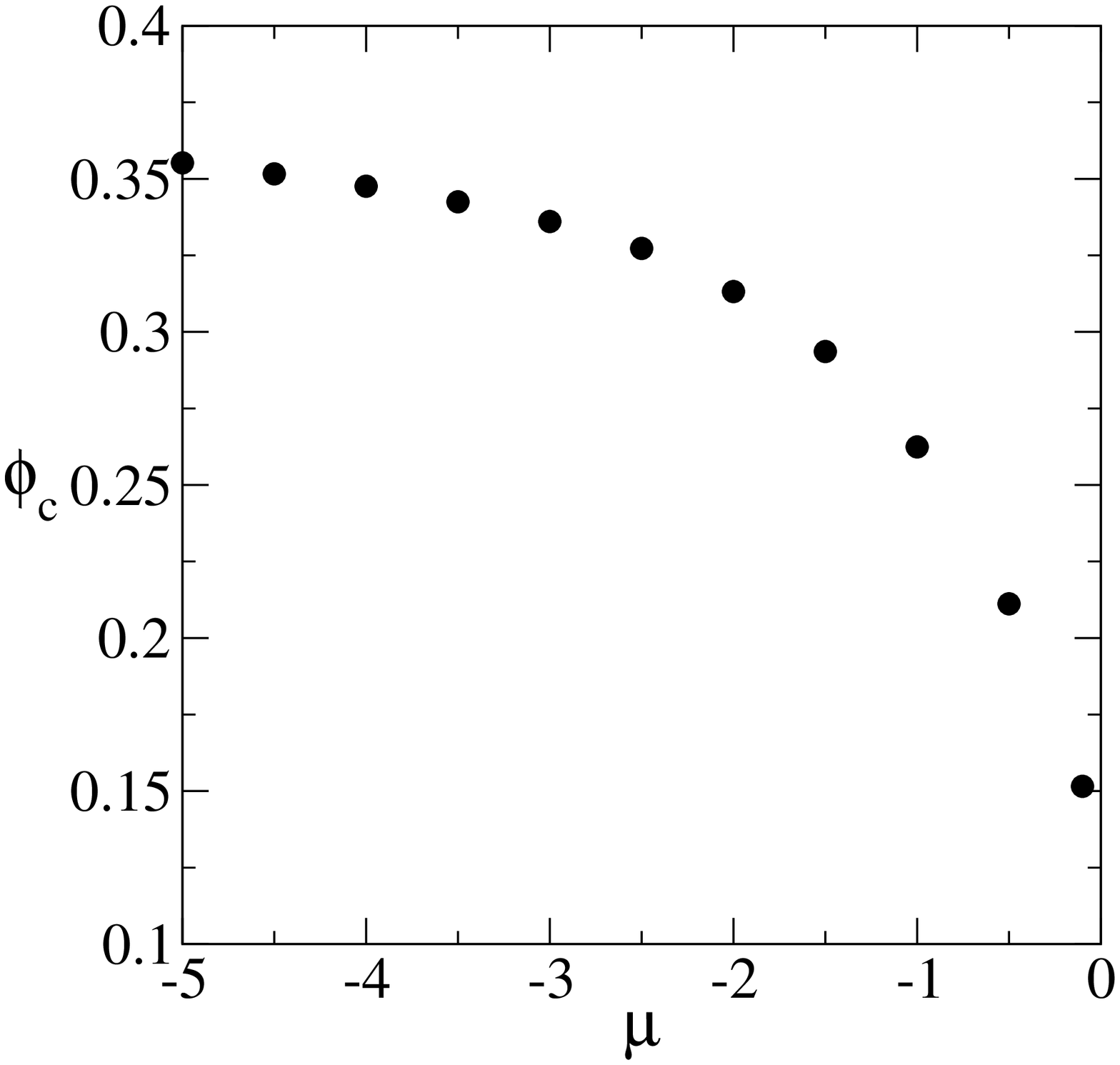,width=8.5cm,angle=270}
\end{center}
\par
FIG. 13 \vspace{5cm}
\end{figure}

\newpage

\begin{figure}[p]
\begin{center}
\epsfig{file=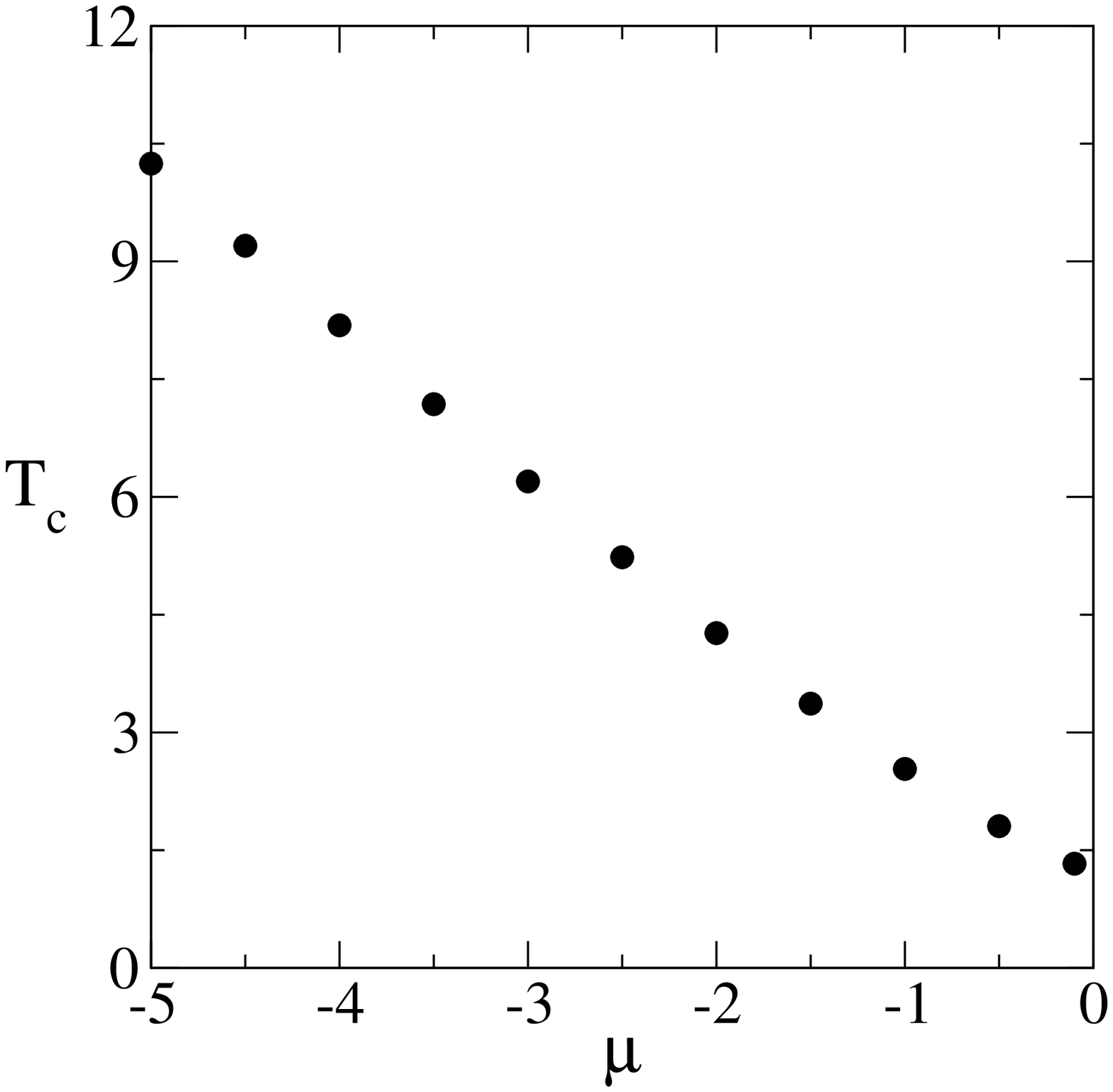,width=8.5cm,angle=270}
\end{center}
\par
FIG. 14 \vspace{5cm}
\end{figure}

\end{document}